\documentclass{pazhb_eng} 

\usepackage{graphicx} 
\usepackage{amsmath} 
\usepackage{epsfig,graphics} 
\usepackage{rotating} 
\usepackage{ulem}

\usepackage{epstopdf}
\usepackage{pdfpages}

\usepackage{longtable}
\usepackage{pdflscape}

\usepackage{hyperref}

\usepackage{comment}
\usepackage{ulem}
\usepackage{color}

\newcommand {\be}{\begin{equation}} 
\newcommand {\ee}{\end{equation}}

\def\red#1{\textcolor{red}{#1}}
\def\БР#1{\red{\textbf{#1}}}

\newcommand{\zspec}{z_{\mbox{\scriptsize spec}}}
\newcommand{\zphot}{z_{\mbox{\scriptsize phot}}}

\newcommand\blfootnote[1]{
	\begingroup
	\renewcommand\thefootnote{}\footnote{#1}
	\addtocounter{footnote}{-1}
	\endgroup
	}

\def\2MASS{\textit{2MASS}}
\def\WISE{\textit{WISE}}
\def\3XMMDR4{\textit{3XMM-DR4}}
\def\SDSS{\textit{SDSS}}

\def\EAZY{\textit{EAZY}}
\def\SigmaDZ{$\sigma_{\Delta z/(1+\zspec)} $}

\def\XMM{\textit{XMM}-Newton}

\newcommand\Nsccat{903} %Число источников в каталоге
\newcommand\Nexclu{63} % Число источников не попавших в каталог
\newcommand{\Voutl}{9} % процент выбросов

\begin{document}

\journalinfo{2016}{42}{5}{277}[294]
%\UDK{524.7}

\title{Catalog of Candidates for Quasars at
 3<\mbox{\small z}<5.5 Selected among X-Ray Sources from the
 \3XMMDR4  Survey of the \XMM \ Observatory }

\author{\bf %\hspace{-1.3cm} \copyright
  G.A.~Khorunzhev\email{horge@iki.rssi.ru}\address{1}, R.A.~Burenin\address{1}, A.V.~Mescheryakov\address{1} and S.Yu.~Sazonov\address{1}
  \addresstext{1}{Space Research Institute, Russian Academy of Sciences, Russia
}}
%\addresstext{2}{Московский физико-технический институт}}
%\footnotetext{\it Институт космических исследований Российской академии наук, Москва, Россия}
%\footnote{\it Институт космических исследований Российской академии наук, Москва, Россия}
%\footnote{\it МФТИ, Московский физико-технический институт}

\shortauthor{Khorunzhev \etal}  
\shorttitle{Quasars of \XMM \ at $3< z<5.5$}

\submitted{\today}
   
\begin{abstract}  
  We have compiled a catalog of
 \Nsccat \ candidates for type 1 quasars at redshifts
  $3<z<5.5$ selected among the X-ray sources of the
  serendipitous\  \textit{XMM}-Newton survey presented in the
 \textit{3XMM-DR4} catalog
 (the median X-ray flux is
  $\approx$5$\times$10$^{-15}$~erg\,s$^{-1}$\,cm$^{-2}$
  in the 0.5--2~keV energy band) and located at high Galactic latitudes
 $|b|>20^\circ$ in Sloan Digital Sky Survey (\SDSS) fields with a total area
of about 300~deg$^2$. Photometric \SDSS \ data as well infrared
  \textit{2MASS}\ and \textit{WISE}\ data were used to select
the objects.
 We selected the point sources from the photometric \SDSS \ catalog with a magnitude error $\delta m_{z^\prime}<0.2$ and a color 
  $i^\prime-z^\prime<0.6$ (to first eliminate the M-type stars).
 For the selected sources, we
have calculated the dependences $\chi^2(z)$ for various spectral templates from the library that we compiled for these purposes using the EAZY software.
 Based on these data, we have rejected the objects whose
spectral energy distributions are better described by the templates of stars at
 $z=0$ and obtained a
sample of quasars with photometric redshift estimates $2.75<\zphot<5.5$.
  The selection completeness of known quasars at $\zspec > 3$ 
  in the investigated fields is shown to be about
 80\%. The normalized median
absolute deviation
 ($\Delta z = | \zspec - \zphot |$) is 
  \SigmaDZ$ = 0.07$, while the outlier fraction is
 $\eta = \Voutl \%$, when
  $\Delta z /(1+\zspec)>0.2$. The number of objects per unit area in our sample exceeds the number of quasars
in the spectroscopic SDSS sample at the same redshifts approximately by a factor of
 $1.5$.
 The subsequent
spectroscopic testing of the redshifts of our selected candidates for quasars at 
 $3<z<5.5$ will allow the
purity of this sample to be estimated more accurately.

  \DOI{10.1134/S1063773716050042}


\end{abstract}

\section{Introduction}
Searching for quasars at $z\gtrsim 3$ is one of the most
important elements of studying the growth history of
supermassive black holes and the evolution of mas-
sive galaxies in the Universe.
 To improve the available
constraints on the models of the {X-ray} luminosity
function for quasars at
$z\gtrsim 3$ requires collecting a
large X-ray sample of distant quasars at high redshifts
 \citep[see, e.g.,][]{aird15}.Bright and dis
tant quasars are rare objects, with X-ray surveys of
large area and sufficient depth being needed for their
search.
Photometric and spectroscopic support at
optical wavelengths is required to work with X-ray
sources.
An increasing role of the methods for estimating the photometric redshift to determine z and to
classify the sources is traceable in present-day works
on this subject matter
 \cite{assefetal11,civano12, kalfonzou14}.

In their recent paper, \cite{kalfonzou14} constructed the X-ray luminosity function for quasars at $z>3$ based on a catalog of 209
X-ray sources detected in various sky fields with a total area of $\approx$33~deg$^2$. This catalog consists of two parts:
\textit{С-COSMOS} is a Chandra X-ray survey with an
area of 0.9~deg$^2$, 122 sources with fluxes\footnote{Here and below, all X-ray fluxes and luminosities are given in the $0.5$--$2$~keV energy band.}  $2\times 10^{-16}<F_X<2\times 10^{-15}$
erg\,s$^{-1}$\,cm$^{-2}$ \citep{elvis09} and \textit{ChaMP}  is a "serendipitous" survey based on CHANDRA pointings with area 33~deg$^2$, 87 sources with fluxes $3\times 10^{-15}<F_X<3\times 10^{-14}$ erg\,s$^{-1}$\,cm$^{-2}$ \citep{kim07}. 
The \textit{C-COSMOS} survey has
deep photometric (using a set of medium-band filters)
and spectroscopic coverage and, for this reason, $\approx$90\% identification completeness of objects at 
 $z>3$. The spectroscopic \textit{ChaMP} program (44 objects) and
various photometric relations (43 objects) were used to measure and estimate the redshifts of
\textit{ChaMP} sources. Thus, objects with firmly established redshifts may be deemed to constitute about 70\% of the sample of quasars at z>3 from Kalfountzou~et~al.(2014).

The observations of numerous astrophysical objects with the
\textit{XMM}-Newton X-ray telescope accumulated for the last 15 years collectively represent
a unique "serendipitous"\ X-ray sky survey \citep{watson09} with a total area of about
 800~deg$^2$ and a sensitivity of
$\approx 5\times 10^{-15}$~erg\,s$^{-1}$\,cm$^{-2}$ \citep[the
\textit{3XMM-DR4} version; 
][]{watson09,watsonXMM13}. It is hoped
that a sample of quasars at $z>3$ selected by their
X-ray emission that exceeds the sample of
\cite{kalfonzou14} by several times will be obtained on
the basis of this survey. This is the goal of our work.

Photometric and spectroscopic support is required
to work with the X-ray sources from the
 \textit{3XMM-DR4} catalog. The solution of the problem is facilitated 
by publicly accessible surveys: the optical photometric and spectroscopic 
Sloan Digital Sky Surveys
\citep[\SDSS,][]{alam15,aiharadr811,eisensteinsdssIII11} with areas of
$\approx$14000 and $\approx$10000 deg$^2$, respectively; the infrared photometric
 \textit{2MASS} \citep{cutri03} and
\textit{WISE} \citep{wright10} all-sky surveys.

The spectra were taken and the redshifts were
measured for a large number of distant quasars as part
of the spectroscopic SDSS programs.
Therefore, it
is natural to use the spectroscopic sample of SDSS
objects (Alam~\etal ~2015) as the first step: $\simeq$60000
quasars with zspec > 3 on $\simeq$ 10000 deg$^2$. However,
the spectroscopic SDSS is based on a fairly complex
method for the selection of quasar candidates whose
completeness and purity depend on many factors: the
redshift, the morphological properties of objects, the
quality of photometry, etc. \citep{ross13}. The
purity of the samples of distant quasar candidates for
the SDSS does not exceed 50\% \citep{ross12}.
At the same time, it is unclear how the completeness
and purity of the selection of candidates for the spec-
troscopic SDSS program behave as a function of the
X-ray flux.

The technique of comparing the spectral energy
distribution of a source in the optical and infrared
bands with the template spectra of quasars, stars,
and galaxies can be used for the selection of quasar
candidates at $z>3$ among the X-ray sources of the \textit{3XMM-DR4} survey.
Such a technique not only gives
the probability that the object is a quasar but also
allows its photometric redshift estimate
 $\zphot$ to be
obtained. In the case of quasars at high redshifts,
$z>3$, the reliability of their redshift estimates based
on optical data increases, because the jump produced
by the bright 
 Ly$\alpha$ line and the
Ly$\alpha$ forest in the quasar
spectrum fall into the optical band.
The addition of infrared \WISE \ photometry increases the purity of
the sample of distant quasar candidates and gives an
advantage over the selection methods based only on
\SDSS photometry.

In this paper, we present the sample of candidates
for quasars at $z>3$ obtained by searching for such
objects among the X-ray sources of the serendipitous
 \textit{XMM}-Newton survey,
\textit{3XMM-DR4}, using \SDSS ,
\2MASS \ and \WISE \ data. The properties of this
sample are discussed. Subsequently, we are going
to use this sample to construct the X-ray luminosity
function for quasars $z>3$.  Here, we use the
following cosmological parameters:
 $\Omega_0$=0.3, $\Lambda_0$=0.7,
$H_0$=70~km\,s$^{-1}$\,Mpc$^{-2}$.

\section{THE SAMPLE OF X-RAY SOURCES WITH
OPTICAL AND INFRARED PHOTOMETRY}
\label{sec:vyborka}

In our work, we used data from the
\textit{3XMM-DR4}\footnote{\url{http://heasarc.gsfc.nasa.gov/W3Browse/xmm-newton/xmmssc.html}}
\citep{watsonXMM13}. Our sample consists of 129541 point
(\texttt{SC\_Extent=0}) X-ray sources
at Galactic latitudes |b|>20$^\circ$ and has an area of
 $\approx$300~deg$^2$, which is determined by the area of the
overlap between the
\textit{3XMM-DR4} survey and the
\SDSS ; the area is calculated below. We work with
the 0.5--2~keV X-ray flux 
(the sum of columns \texttt{SC\_EP2\_FLUX} and \texttt{SC\_EP3\_FLUX}
in the \textit{3XMM-DR4} catalog).

We cross-correlated the catalog of X-ray sources
with the data from the photometric (data release 10)
and spectroscopic (data release 12) \SDSS \
\citep{aiharadr811,ahn12,alam15,eisensteinsdssIII11}. Within the
2$\sigma$ X-ray source position error
 (if it was less than 3\arcsec, then within 3\arcsec) we searched for all optical \SDSS \ sources. As a result, we produced a catalog of 64714 X-ray sources
that have an optical counterpart among the \SDSS \
objects.
At the same time, 2489 (4\%) X-ray sources
have more than one optical counterpart. In the case of
an ambiguous correspondence, we estimated
 $\zphot$ for all the possible optical counterparts of a given X-ray
source. It turns out that there are spectroscopic
data for 12823 ($\approx20\%$ of the sample) X-ray sources
in the \SDSS.
 We deemed the spectroscopic redshift
measurement reliable at the flag \texttt{zWarning=0}. It turns
out that 390 sources in the
 \textit{3XMM-DR4} catalog have ${\zspec>3}$ and
\texttt{zWarning=0}; 280 of them have a magnitude measurement error
 $\delta m_{z^\prime}<0.2$ (see below).

We additionally made a cross-correlation with the
catalog of quasars from \cite{flesch15a}. This catalog 
contains the spectroscopically confirmed X-ray
quasars the data on which were published before
January 25, 2015, and the sample of reliable (the
probability that >99\% will turn out to be a quasar) 
photometric candidates from
\cite{bovy12,richards09}.
We added 49 objects with
$z>3$ from the catalog of \cite{flesch15a}, five of them
with photometric $z$ estimates\citep{richards09}. 
As a result, we obtained a spectroscopic sample
of 329 quasars with redshifts greater than 3 (280 from the spectroscopic SDSS sample and 49 from
Flesch (2015)). This sample is required to check the
completeness of our sample of quasars (see below).

We searched for infrared counterparts in the
\textit{2MASS} \citep{cutri03} and \WISE \ \citep{wright10} catalogs. 
The search was made within
6~arcsec of the X-ray source position. If we found no
infrared counterpart for an optical source, then we set
 $7\sigma$ and $5\sigma$ limits for the
\textit{2MASS}  photometric
 $J$, $H$, $K\!s$ bands and \textit{WISE} photometric
 $w1$, $w2$, $w3$, $w4$ bands, respectively, using the photometric data for
adjacent sources. For the
 \2MASS\ and \WISE\ sources
that were not detected in all bands,
we used the upper
limits from the corresponding catalogs.
We assumed that the
 \2MASS\ sources always corresponded to
the sources with \WISE\ photometry.
When a \WISE\ source was offset from an
 \SDSS\ source by more than
2 arcsec, then we additionally considered the case
where the source had only \SDSS\ photometry when
calculating $\zphot$. In this case, the upper limits in the
 \2MASS\ and \WISE \ filters were taken to be equal to
the \WISE/\2MASS\ magnitudes of the controversial
source.

Thus, we have a flux measurement or an upper
limit on the flux in 12 broad bands ($u^\prime$, $g^\prime$, $r^\prime$, $i^\prime$,
$z^\prime$, $J$, $H$, $K\!s$, $w1$, $w2$, $w3$, $w4$) for the X-ray sources.
To calculate the fluxes in all photometric bands, we
used the magnitudes designed to measure the fluxes
from point sources, which the distant quasars are. In
addition, the search was made only among the point \SDSS\ sources.

There is no photometry for bright stars in the
 \SDSS. To take into account their influence on
our quasar selection method, we searched for the
counterparts of \textit{XMM}-Newton X-ray sources within
 6~arcsec in the Guide Star Catalog
 \citep[\textit{GSC},][]{lasker90}. This catalog was produced by processing survey plates from the Palomar (northern hemisphere), Siding Spring (southern hemisphere),
and several more observatories. The \textit{GSC} contains
  $\sim$19 million sources with magnitudes from 6 to 16
(note that the sources brighter than magnitude 15
are overexposed in the SDSS), some of which were
astrometrically classified as stars.

\begin{figure}
\centering   

\includegraphics[width=0.99\linewidth]{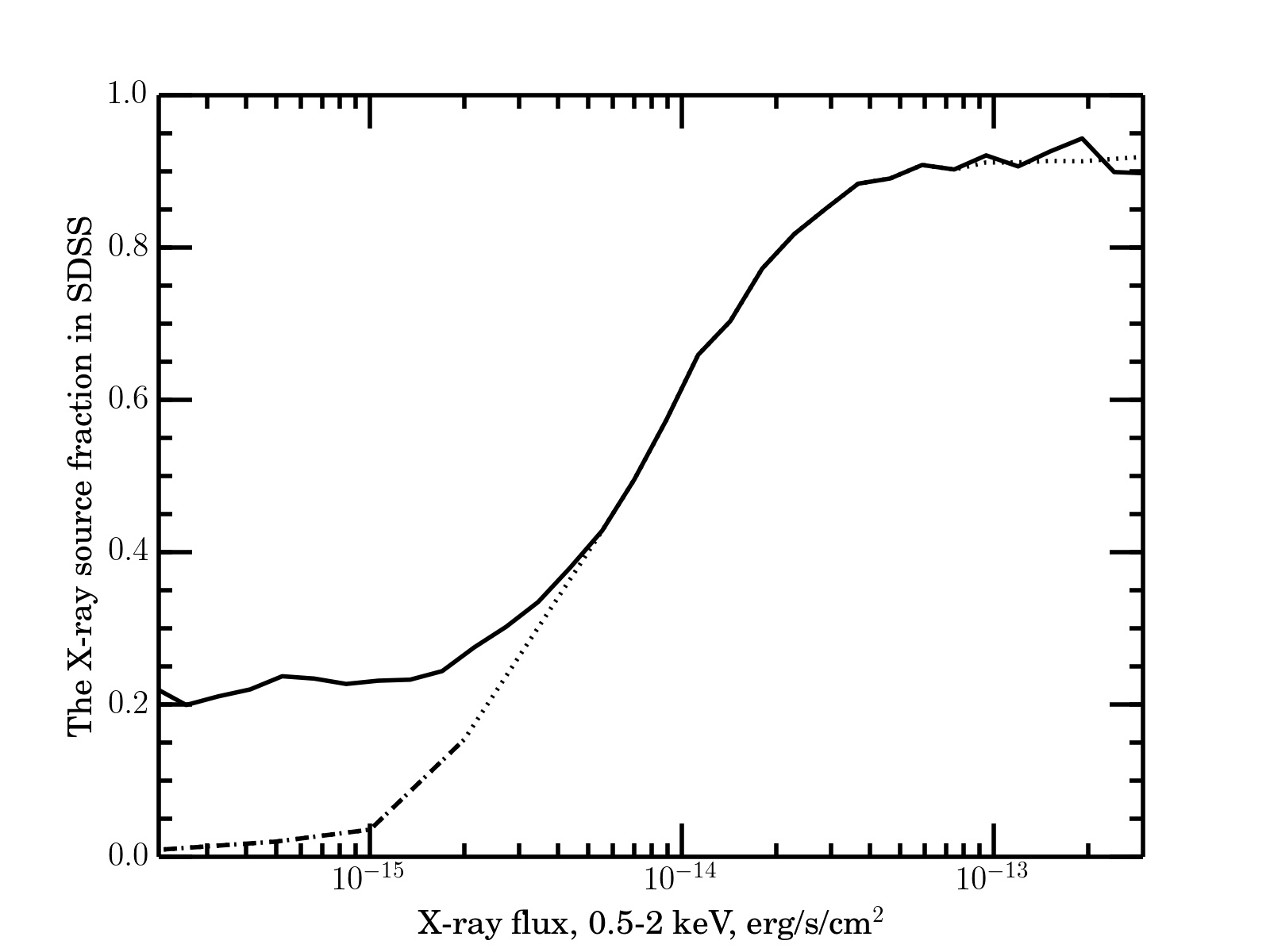}
\caption{The solid line indicates the fraction of X-ray sources in the \3XMMDR4
catalog that were detected in the photometric \SDSS \
  ($\delta m_{z^\prime} <0.2$) or that have a counterpart in the \textit{GSC}
   as a function of the X-ray flux.
The dashed line (at low X-ray fluxes) indicates the fraction of quasars at
 $z>3$ from \textit{COSMOS} \citep{kalfonzou14} detected in the photometric \SDSS. The dots indicate the detection completeness function for X-ray quasars at $z>3$ in the \SDSS\ composed from these
two dependences and used below versus the X-ray flux.
 }
\label{fig:DR12fluxtotal} 
\end{figure}

The fraction of X-ray sources in the \3XMMDR4\ catalog for which there are optical counterparts in the photometric \SDSS\ and the
\textit{GSC}  drops with decreasing X-ray flux from
$\approx 80\%$ at $F_X\approx 10^{-14}$~erg\,s$^{-1}$\,cm$^{-2}$ to
$\approx 20\%$ at $F_X\approx 10^{-15}$~erg\,s$^{-1}$\,cm$^{-2}$.
At lower X-ray
fluxes, the fraction of nearby galaxies ($z$ < 1) in
the sample of X-ray sources increases \citep{lehmer12} ;
therefore, no further reduction in the
detection completeness of objects in the SDSS
occurs.\\

The quasars at $z>3$ constitute a small fraction
of all X-ray sources, and the fraction of objects with
optical counterparts for them can differ from that for
all X-ray sources. This assumption can be tested
using the representative sample of quasars with
reliably measured $z>3$ from the deep С-COSMOS
X-ray survey that was
talked about above. Out of 122 objects in this sample,
32 have spectroscopic redshifts
$\zspec$, 75 have reliable
photometric redshifts $\zphot$ (measured with a set of
medium-band filters), and 15 have no optical counterpart to
$m_{i^\prime}\approx 28$ and, therefore, are also deemed
probable candidates for quasars at $z>3$ \citep{kalfonzou14,civano11}.
Figure \ref{fig:DR12fluxtotal} shows how
the fraction of objects from this sample detected in
the photometric SDSS changes with X-ray flux. The
fraction of objects with optical counterparts detected
in the SDSS among the quasars at $z > 3$ turns out
to be lower than that for all X-ray sources at fluxes
below 2$\times$10$^{-15}$erg s$^{-1}$ cm$^{-2}$ by several times.

Below, when taking into account the influence
of object detection incompleteness in the photometric 
\SDSS\ on our final sample of candidates for
quasars at $z > 3$, we use an approximate smooth
photometric completeness function of the X-ray flux
composed from the dependence for the detection
completeness of quasars at $z > 3$ at fluxes below
2$\times$10$^{-15}$~erg~s$^{-1}$~cm$^{-2}$
and the corresponding
dependence for all X-ray sources at fluxes above
5$\times$10$^{-15}$~erg~s$^{-1}$~cm$^{-2}$ (see Fig. \ref{fig:DR12fluxtotal}).

%\section{Метод измерения фотометрического красного смещения с применением программы \EAZY}
\section{PHOTOMETRIC DATA FITTING BY
SPECTRAL TEMPLATES}

In our work, to calculate the 
 $\chi^2$ при values when fitting
the spectral energy distribution in the optical and infrared 
bands by various spectral templates and to obtain the photometric redshift estimates, 
we used the \EAZY\ \citep{brammerdokkum08}.
This software steps through a grid of redshifts and at each
$z$ finds
the best-fitting synthetic template by minimizing the
 $\chi^2$ value:
\begin{equation}
  \mbox{$\chi^2$} = \sum_{j=1}^{N_{filt}} \frac{(T_{z,i,j}-F_j)^2}{(\delta F_j)^2} \ ,
  % \mbox{$\chi^2$} = \sum_{j=1}^{N_{filt}} \Bigr( \frac{T_{z,i,j}-F_j}{\delta F} \Bigl)^2 \ ,
  \label{for:eazy}
\end{equation}
where $N_{filt}$ is the number of photometric data points
(filters), $T_{z,i,j}$ is the synthetic flux of template
 $i$ at redshift $z$ in filter $j$,
$F_j$ --- is the observed flux in filter
 $j$ and $\delta F_j$ --- is the flux error that takes into account the
photometric error in filter $j$.

The \EAZY\ software implements an algorithm that
allows one to seek a superposition of several spectral
templates, for example, the spectra of a galaxy and
a quasar. However, we use only a one-template fit
in our work, because we search for bright quasars
whose emission dominates over the emission from
the host galaxy. The quasars are characterized by
a large scatter of spectra and strong variability
\citep[see e.g.,][]{richards06}; therefore, it is necessary to
have a sufficiently complete template library.
\\

For each object, we made separate iterations for
the libraries of quasar and star templates. When fitting by the quasar templates, 
we corrected the \SDSS \ photometry for extinction in our Galaxy using its
values for each object from the photometric \SDSS\
catalog. For the star templates, the \SDSS\ photometry 
was not corrected for extinction in our Galaxy.
The best value of $\zphot$ was sought on the grid of redshifts
$[0.01,7]$ with a 0.01 step by finding the best-fitting
template from the library of quasars according to the $\chi^2$ test.

We also took into account the intergalactic ab-
sorption by neutral hydrogen \citep{madau95}. he
intergalactic absorption allows for the influence of the
 Ly$\alpha$ forest on the spectral energy distribution of a
quasar. At redshifts $z>3$, Ly$\alpha$ forest absorbs
a substantial fraction of light on the blue side of the Ly$\alpha$ line,
and this fraction increases with redshift. At
such redshifts, the Ly$\alpha$ line falls into the photometric
$g^\prime$ band, and the presence of a jump in
аличие скачка поглощения Ly$\alpha$-forest absorption can already be determined from
 $u^\prime-g^\prime$ color.  At redshifts above
$z\approx 3.5$, the Ly$\alpha$ line moves into the $r^\prime$ band,
and the presence of an absorption
jump is determined from the
$g^\prime-r^\prime$. This feature
in the spectra of quasars at redshifts
 $z>3$ allows one
to increase the identification reliability of such objects
and to obtain more reliable photometric estimates of
their redshifts (see below).

\subsection{The Template Library}
\label{sec:eazylibrarysed}

\begin{figure*}
\centering   
\vskip -2.3cm
\includegraphics[width=0.7\linewidth]{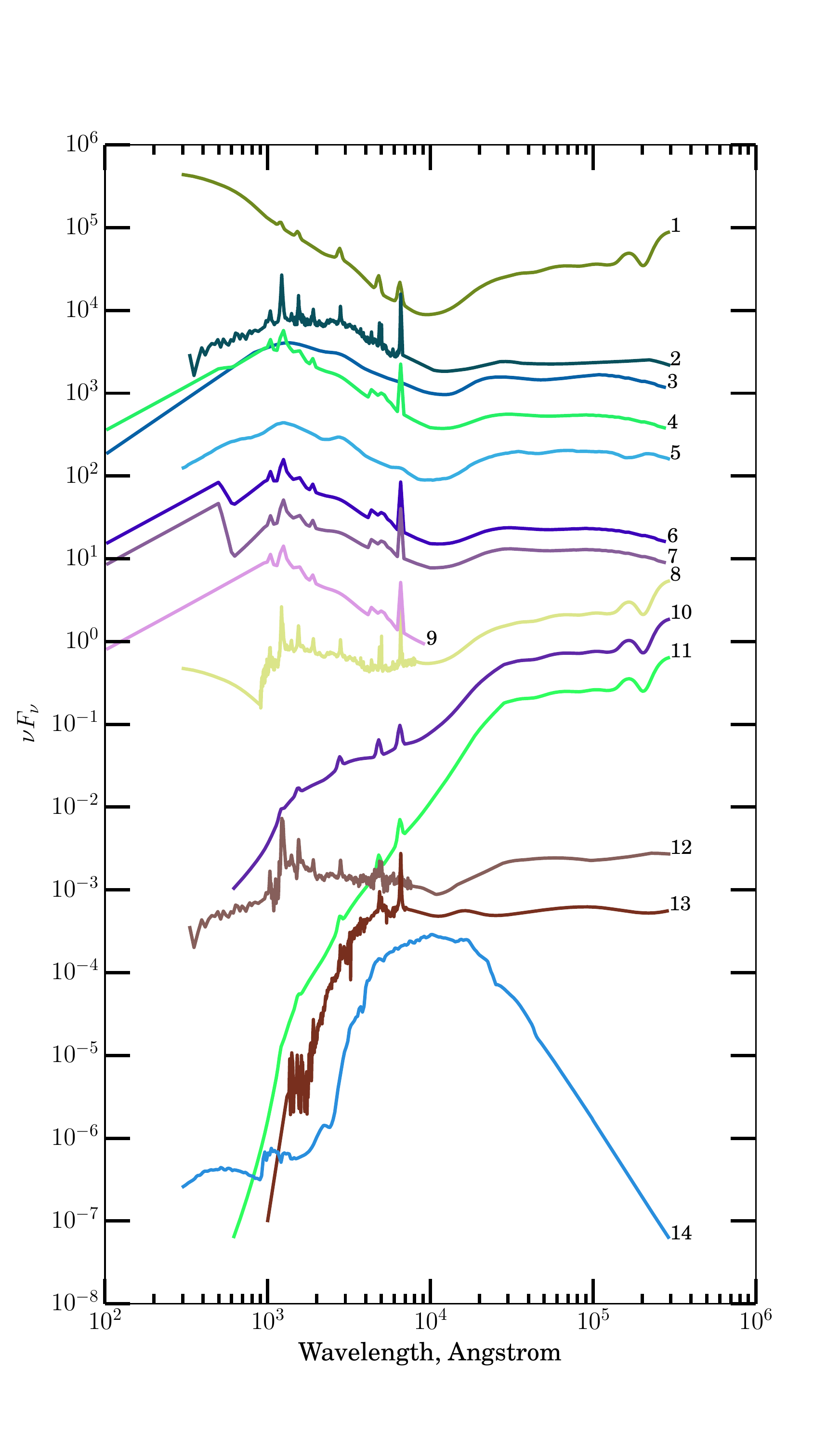} 
\vskip -1.cm
\caption{Templates from the library of quasars in relative units
  $\nu F_\nu$: 1 --- the average spectrum of type 1 quasars
  \citep{assefetal10}, 2 --- the average spectrum of \SDSS \ quasars 
  \citep{brammerdokkum08}, 3 --- the average spectrum of luminous type 1
quasars \citep{krawczyketal13}, 4 --- the average spectrum of type 1 quasars
 \citep{krawczyketal13} supplemented by emission
lines with extinction $E(B-V)=0.01$, 5 ---the average spectrum of type 1 quasars (variant 2) 
 \citep{assefetal10}, 6 --- the
average spectrum of type 1 quasars
 \citep{krawczyketal13}, supplemented by emission lines with extinction
 $E(B-V)=0.05$, 7 --- the average spectrum of type 1 quasars
 \citep{krawczyketal13}, supplemented by emission lines with extinction
 $E(B-V)=0.1$, 8 --- the template of a type 1 quasar extended to the infrared 
 \cite{vandenberk01}, 9 --- the average
spectrum of type 1 quasars \citep{krawczyketal13} truncated to fit only the optical part,
  аппроксимации только оптической части, 10 --- template~1
 \citep{assefetal10} with extinction $E(B-V)=0.5$ , 11 --- template~1
 \citep{assefetal10} with extinction $E(B-V)=1.0$, 12 --- the composite quasar
template from the spectra of the deep \textit{VIMOS-VLT} survey
   \citep{gavignaud06,ilbert06}, 13 ---t he spectrum of an obscured type 2 quasar
 \cite{polletta07}, 14 --- the spectrum of an elliptical E0 galaxy
 \citep{assefetal10}.}
\label{fig:tempqso} 
\end{figure*}

The targets of our search are type 1 quasars
(without any significant intrinsic absorption) at $z>3$. 
We used the spectral templates of quasars from
the \EAZY\ \citep{brammerdokkum08} and 
\textit{LePHARE} \citep{ilbert06} libraries as well as from other
libraries
\citep{assefetal10, richards06, krawczyketal13}. The templates that we
modified \citep{krawczyketal13} were added to the library. In these templates, we took into account the
contribution of the quasar's emission lines obtained
from the template
\citep{vandenberk01}. Besides, we extended the average spectrum of \SDSS\
quasars
 \citep{vandenberk01} to the infrared
($\lambda >8000$~$\mathring{A}$) using the template of a type 1 active
galactic nucleus (AGN) \citep{assefetal10} and added
the resulting template to our library.
We also added
the templates of an elliptical galaxy \citep{assefetal10} 
and a type 2 quasar \citep{polletta07}.
As in \citep{dahlen13}, we applied the set of extinctions
 $E(B-V)=0.01$, $0.05$, $0.5$ and $1.0.$ according to the
law of
\citep{calzetti00} using tabulated data from
the \textit{LePHARE} software package
 \citep{ilbert06} to the templates.

We tested the template library on the spectroscopic
\SDSS-DR12 \ sample and eliminated those
templates that more often erroneously classified a
nearby object as an object at $z>3$ than correctly
classified distant quasars from the template library.
Thereafter, 14 templates remained, which were 
subsequently used to search for quasars at $z>3$. These
templates are presented in Fig.~\ref{fig:tempqso}.

The galaxy and obscured quasar templates are
needed only to eliminate the nearby objects whose 
accurate redshift is of no interest to us. In this case, no
additional templates of galaxies (star-forming ones,
with a different contribution of the active nucleus,
etc.) are required to be applied, because this will
increase the number of outliers and degrade the
completeness and purity of the final sample of quasars at
$z>3$ \citep{simm15,salvato09}.

For the templates of stars, we used the library of
spectra from \cite{pickles98}. It includes the spectra
of spectral types $O$,
$A$, $B$, $F$, $G$, $K$, $M$. For some
of the templates, there is a breakdown into luminosity
classes: i --- supergiants, ii --- luminous giants, iii ---
giants, iv --- subgiants, v --- main-sequence stars. The
library spectra were constructed only to a wavelength
of 2.5~$\mu$m \cite{pickles98}. Since the \WISE\ filters
are longer-wavelength ones, each spectrum from this
library was extrapolated by Planck’s law for black-
body radiation to a wavelength of 30~$\mu$m. Such assumption
works well for single main-sequence stars hotter
than M0.

\begin{figure*}
\centering   
\includegraphics[width=0.45\linewidth]{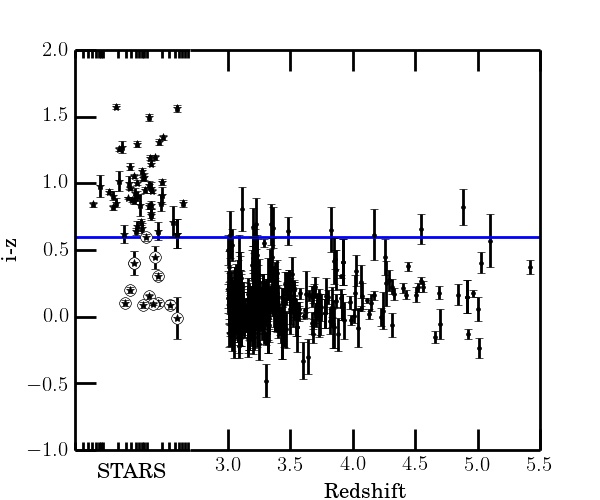}
\includegraphics[width=0.45\linewidth]{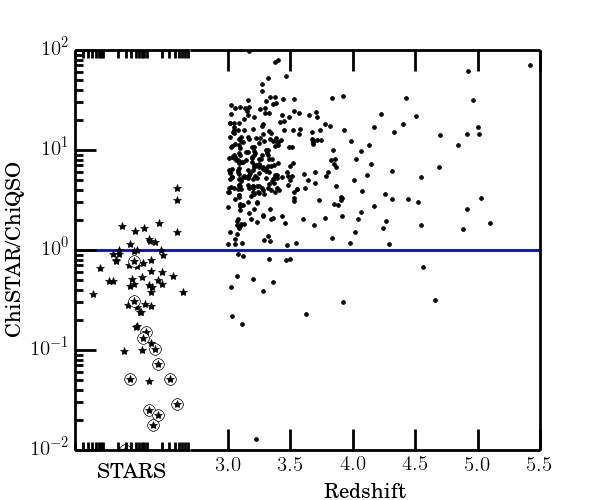}
\caption{Criteria for selection by the $i^\prime-z^\prime$ \SDSS (left) and
  $\chi_{star}^2/\chi_{qso}^2>1$ (right) versus redshift. The solid horizontal
  lines indicate the boundaries of the selection criteria. The dots indicate 329 
  quasars with $\zspec>3$.
  The symbol $\star$ indicates 66 spectroscopically confirmed stars from \SDSS \ DR12
  that have a minimum of the $\chi^2(\zphot)$ distribution at
  $\zphot >3$.  The circles
  mark the stars with the same color $i^\prime-z^\prime$<0.6 as that for the quasars. }
\label{fig:Zgt3criterion} 
\end{figure*}

\subsection{Criteria for the Selection of Quasar Candidates $3<z<5.5$}
\label{sec:eazypreliminary}
Since the quasars are star-like objects, we 
consider only the point SDSS sources when searching
for distant quasars. In this case, the contamination
of the sample by M-type stars presents a significant
problem. Many of them are detected in the \XMM \ 
X-ray survey owing to their coronal activity.
During flares, the ratio of the X-ray and optical/near-IR 
fluxes for M dwarfs can be the same as that for
quasars.

The SDSS sensitivity in the blue filters is in-
sufficient to separate distant quasars with
$i^\prime \gtrsim$20.5 from stars. M dwarfs can be separated from distant
quasars only by the
$i^\prime-z^\prime$ \citep{richards02, wuhao12, skrzypek15}: the stars
have $i^\prime-z^\prime$>0.8, while the quasars have $i^\prime-z^\prime$<0.4. 
The color for the quasars $i^\prime-z^\prime<0.4$ remains
approximately constant up to a redshift of $\approx$5.5, 
until the Ly$\alpha$ \ line passes from $i^\prime$ to $z^\prime$. 
The color of the
quasars then becomes $i^\prime-z^\prime\approx 2$. The quasars at
$z\gtrsim 5.5$ have reliable photometry only in one \SDSS \ band, and, 
in that case, the method of determining $\zphot$ by template fitting works poorly. 
Therefore, to increase the purity of our sample of quasar candidates,
we used a constraint on the sensitivity in the $z^\prime$ SDSS band and the $i^\prime-z^\prime$ color:
\begin{equation}
\delta m_{z^\prime}<0.2 \ \& \ i^\prime - z^\prime <0.6 \ ,
\label{eq:CriterionQSOnotM}
\end{equation}
where $\delta m_{z^\prime}<0.2$ is the error in the apparent 
magnitude of a point source in the $z^\prime$ \SDSS \ filter. 

The constraint
$\delta m_{z^\prime}$<0.2 help to reduce 
the selection effects in searching for quasars. 
The spectral
energy distribution of quasars rises from
$\lambda \approx$1~$\mu$m to $Ly\alpha$. 
The relations between the threshold sensitivities
of the filters in the \SDSS\ are such that by imposing
the constraint (2), we guarantee the detection com-
pleteness of type 1 quasars not only in the $z^\prime$ filter but
also in bluer \SDSS\ filters.

After applying the photometric constraints, the
quasar candidates are selected according to the condition:
\begin{equation}
%\begin{split}
  \chi_{star}^2/\chi_{qso}^2>1 \ \& \  \zphot >2.75 
  % (\ z_{spec} > 3 \ \& \ zwarning=0 \ ) \ || \ (z_{phot} >3 \ \& \ \chi_{star}^2/\chi_{qso}^2>1 )\ , 
  \label{eq:Criterionzgt3}
%\end{split}
\end{equation}
where $\chi_{star}^2$ is the smallest $\chi^2$ value for the template
from the library of stars, and $\chi_{qso}^2$ is the primary minimum of the $\chi^2$
distribution for the quasar templates. 

The conditions $\chi_{star}^2/\chi_{qso}^2>1$ and 
$i^\prime-z^\prime$<0.6 eliminate 97\% of the stars among the X-ray sources of the
 \3XMMDR4 \ catalog. This was tested on stars for
which the SDSS spectra are available 
(see~Fig.~\ref{fig:Zgt3criterion}).

There exists a scatter of $\zphot$ relative to $\zspec$.
Hence, some of the objects with $\zspec>3$ will have
$\zphot \lesssim 3$; therefore, we chose the lower limit $\zphot > 2.75$
when compiling the catalog of candidates for
quasars at $z > 3$.

We additionally checked our sample of quasar
candidates to eliminate, where possible, the source
misidentification errors. We excluded 20 objects from
the sample: two stars with the \SDSS\ spectra, several
faint sources against the background of a nearby
bright galaxy, faint sources near a bright star, and
objects with erroneously measured redshifts.

\begin{figure*}
  \begin{center}
    \includegraphics[width=0.6\linewidth]{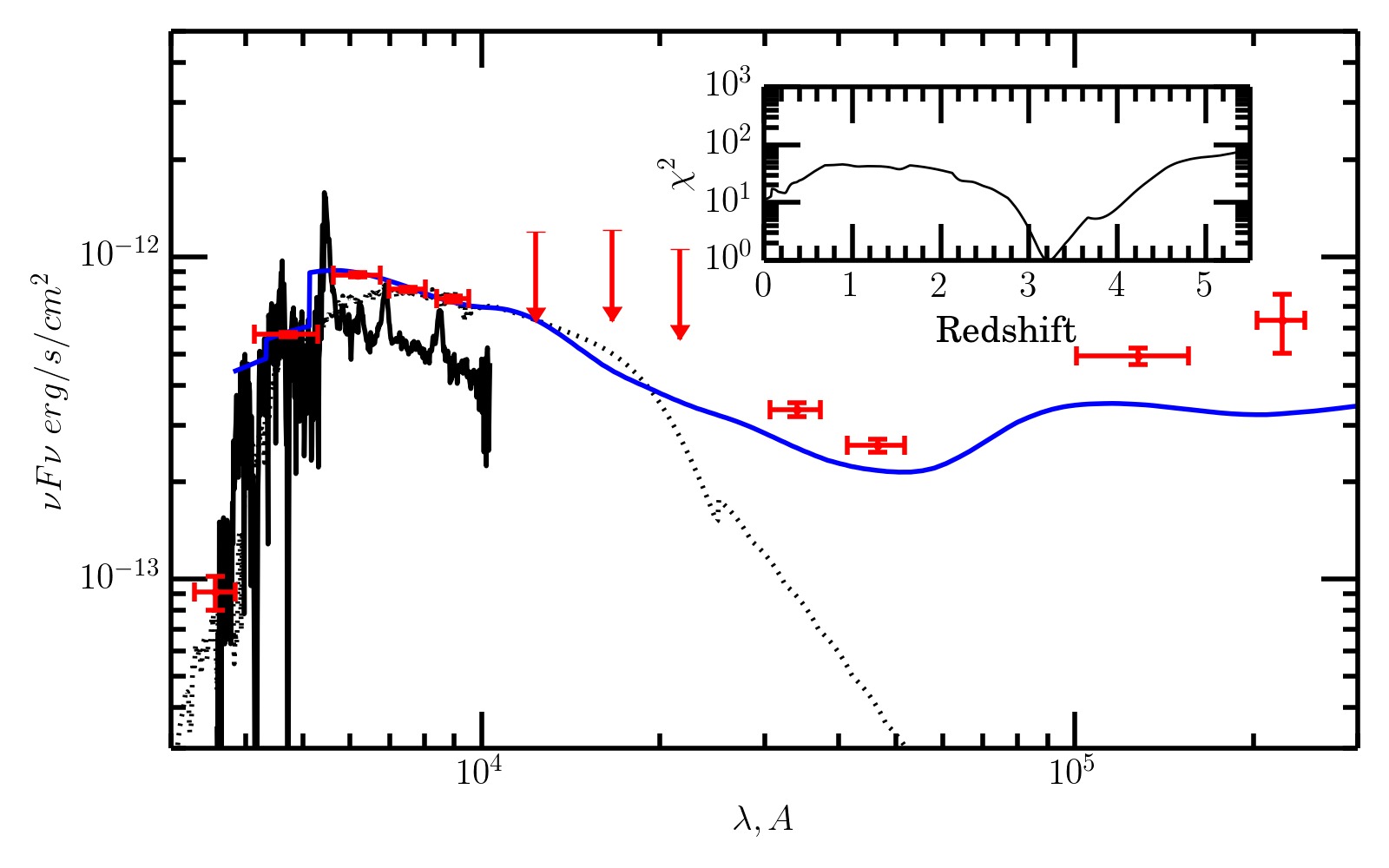}
    \includegraphics[width=0.6\linewidth]{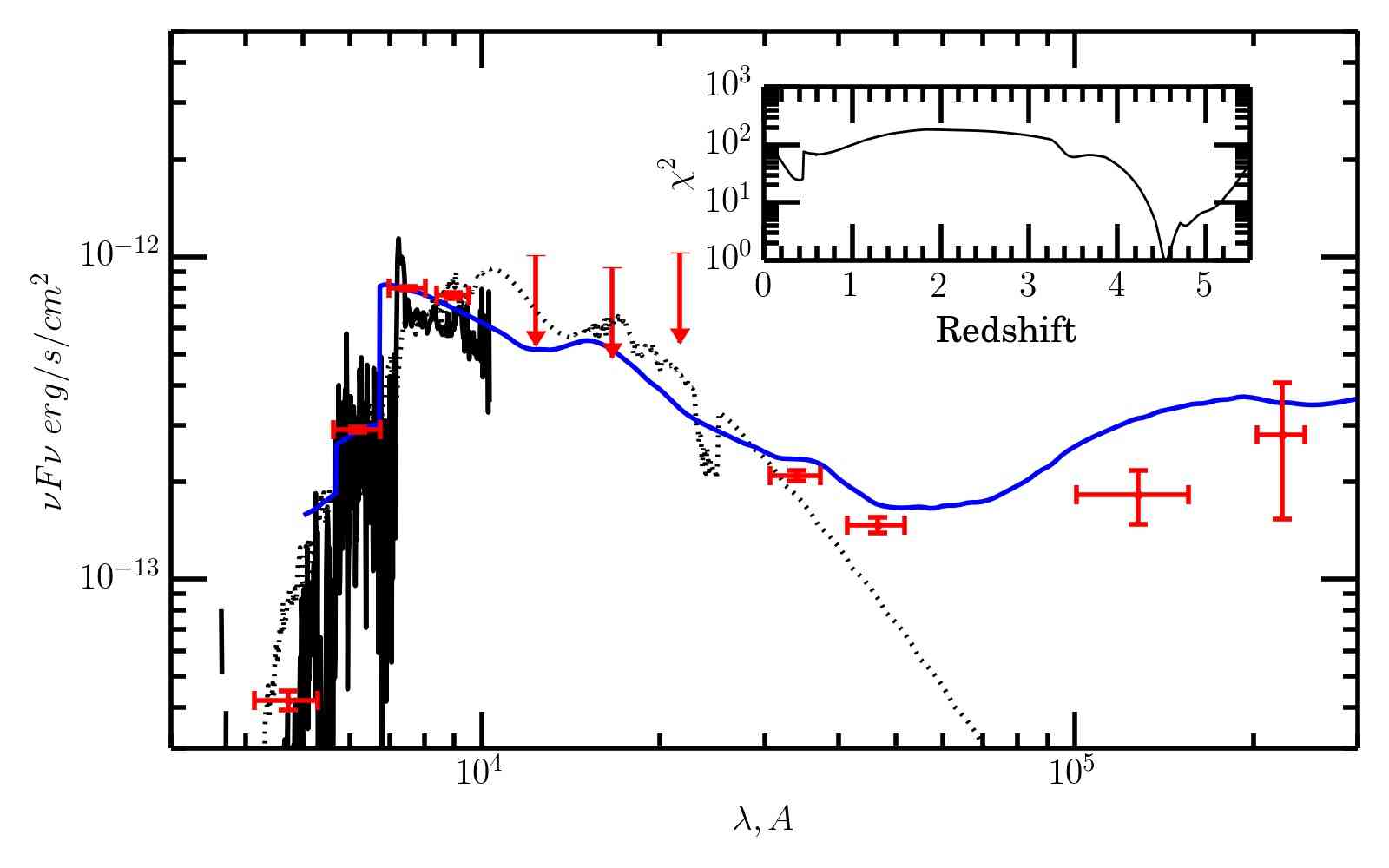}
    \includegraphics[width=0.6\linewidth]{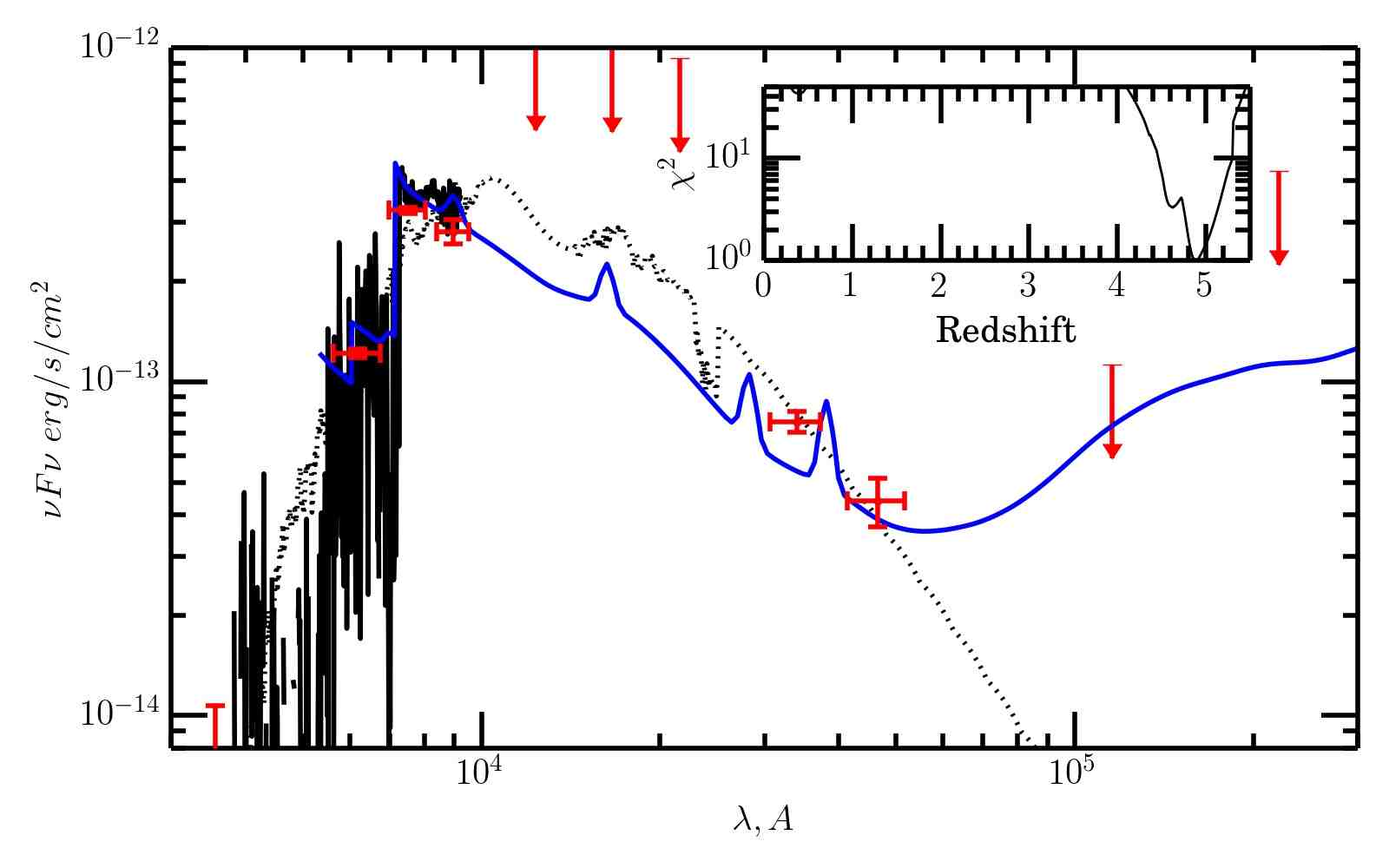}
    \caption{Examples of a satisfactory determination of the photometric redshift.
     Top: \textit{3XMM~J151147.1$+$071406} is a quasar,
      $\zspec =3.481$, $\zphot =3.22$. Middle:
      \textit{3XMM~J001115.2$+$144601}  is a quasar, $\zspec =4.964$,
      $\zphot = 4.54$. Bottom: \textit{3XMM~J004054.6$-$091527}  is a quasar,
      $\zspec =5.020$, $\zphot =4.88$. The \SDSS \ spectrum (black solid line), 
      the photometric data points with their errors (red), 
      the template of a quasar at $\zphot$ (blue solid line), and the template of a star (black dots) 
      are shown. The $\chi^2(z)$ distribution when fitting the photometric data by quasar templates is shown in the insets.}
    \label{fig:exampleGOODSED}
  \end{center}
\end{figure*}

\begin{figure*}
  \begin{center}
    \includegraphics[width=0.45\linewidth]{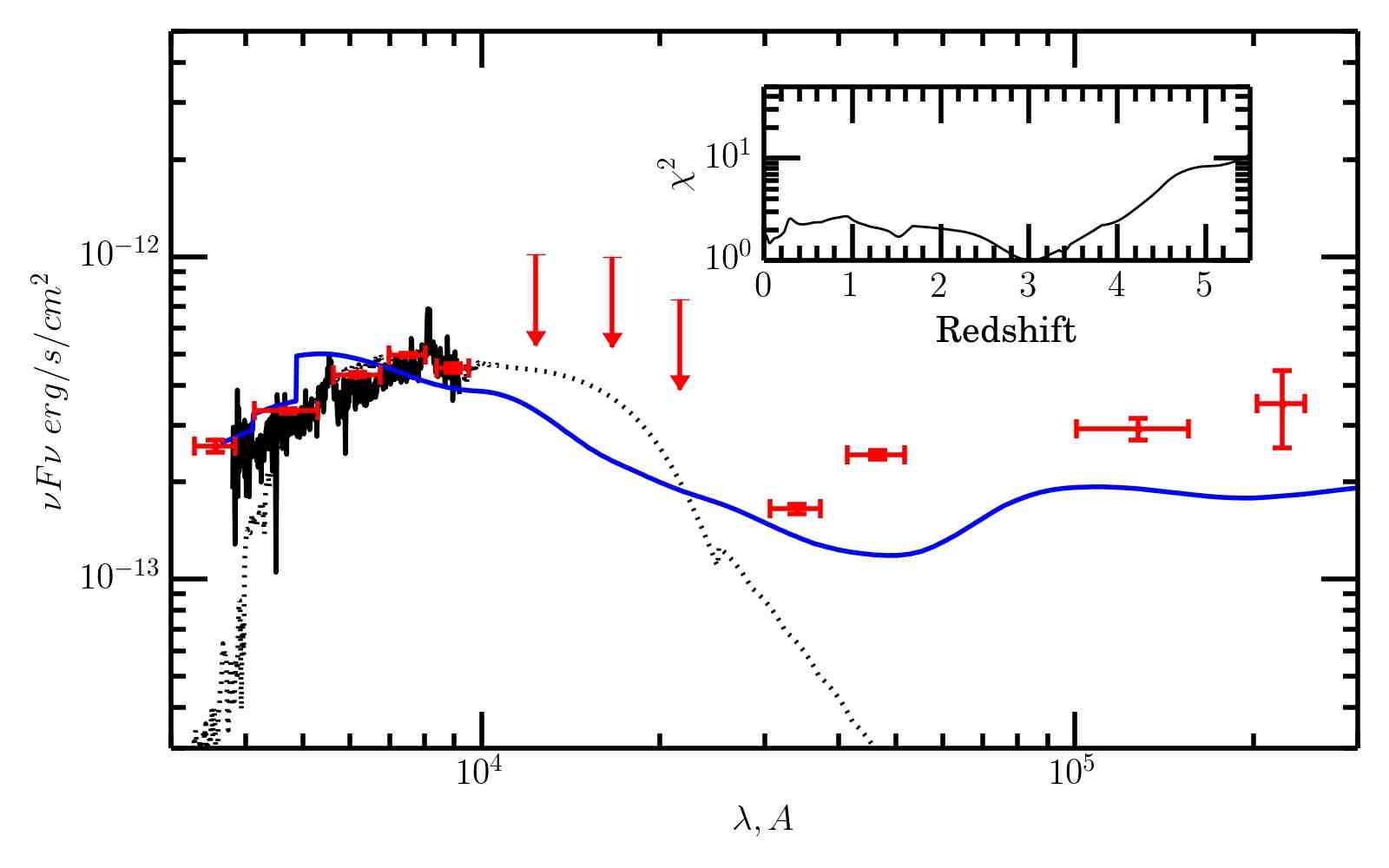}
    \includegraphics[width=0.45\linewidth]{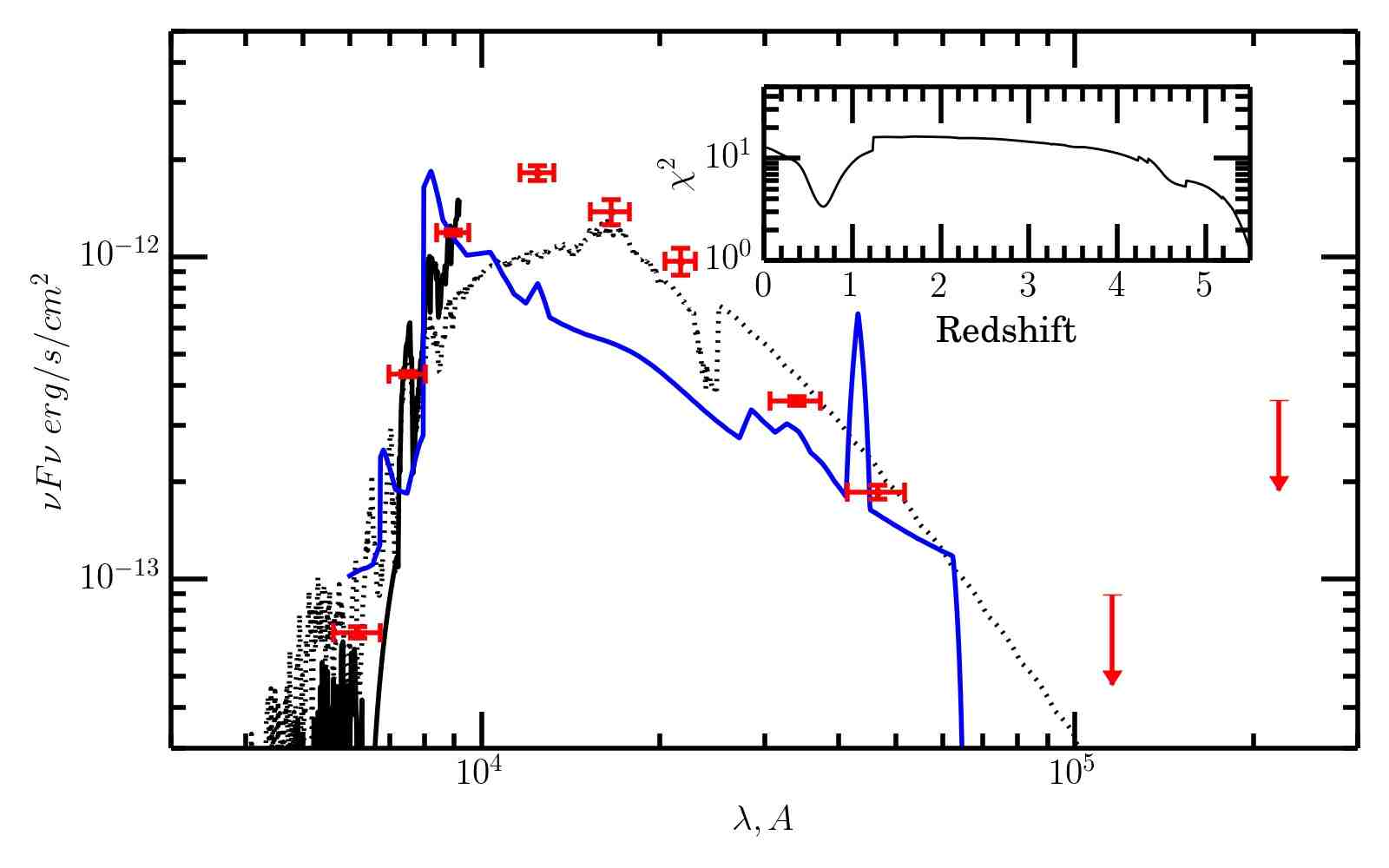}
    \includegraphics[width=0.45\linewidth]{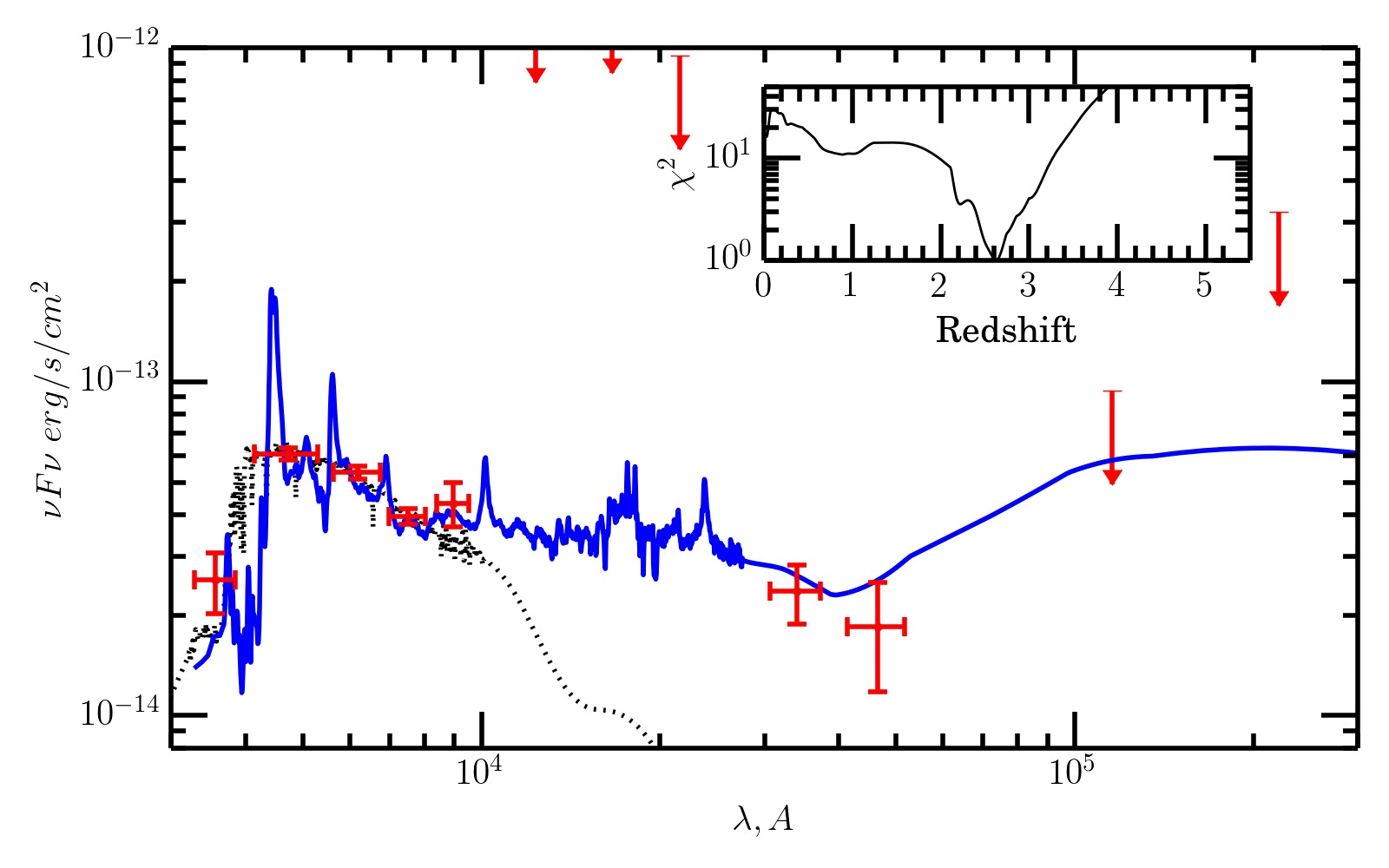}
    \includegraphics[width=0.45\linewidth]{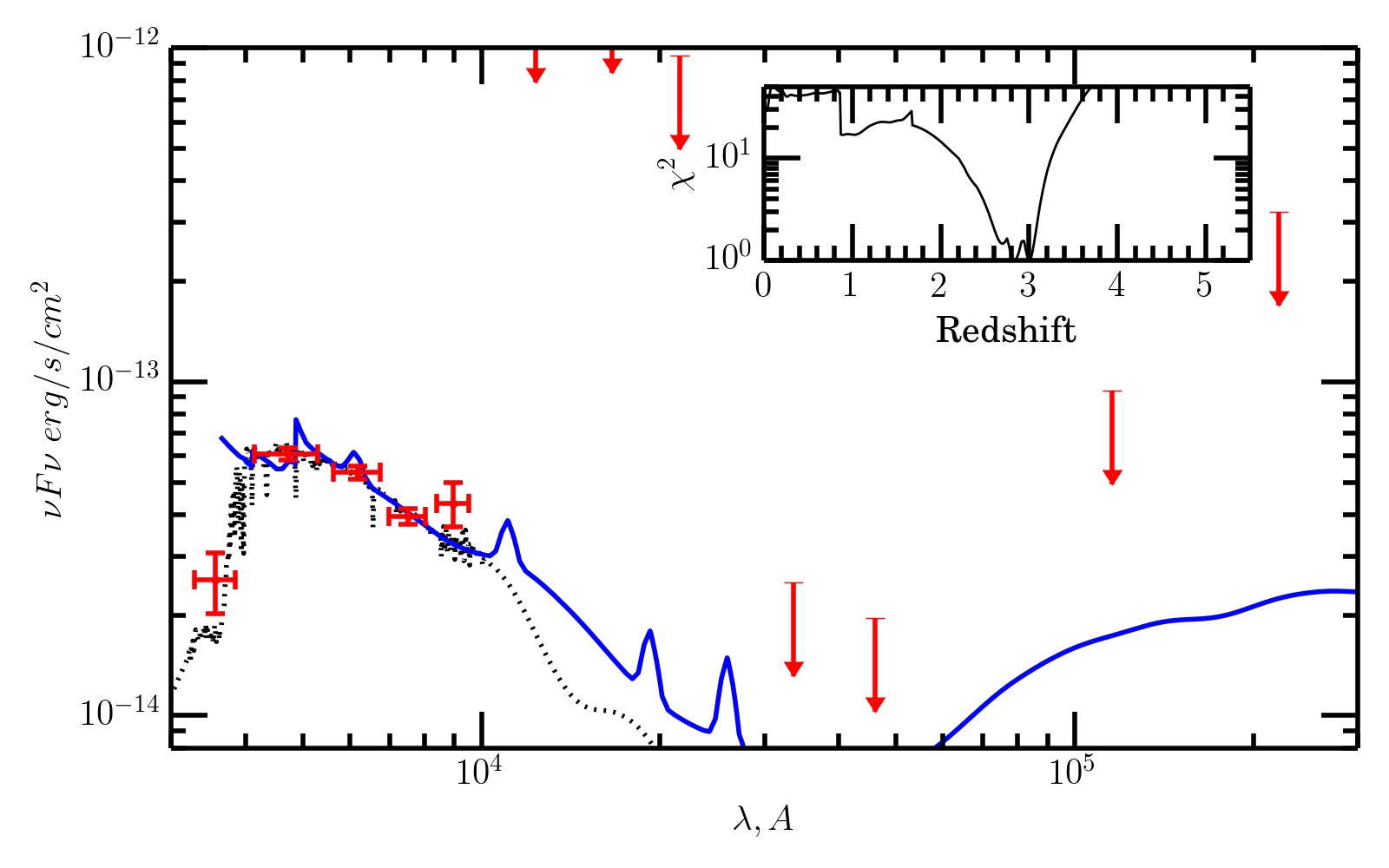}
    \caption{Example of an unreliable estimation of the photometric redshift.
     Upper left: \textit{3XMM~J153322.7$+$324351} is a quasar at
      $\zspec =1.897$. Its photometry is best described by the template 
      of a type 1 quasar with $\zphot$ = 3.06. Upper right:
       \textit{3XMM~J080630.4$+$153241} is an M--type star with
      $\zphot =5.58\pm0.02$. The object was correctly classified as a star and
      was not included in the catalog of quasar candidates, 
      because $i^\prime-z^\prime$>0.6 and $\chi_{qso}^2 >\chi_{star}^2$ for it.  Bottom left and right:
      \textit{3XMM~J094109.9$+$344902} is a quasar at $\zspec =2.643$, 
      for which two photometric redshift estimates were obtained due to its
      controversial association with a \WISE: $\zphot =2.62$ and
      $\zphot =3.00$). The designations are the same as those in Fig. 4. }
    \label{fig:exampleBADSED}
  \end{center}
\end{figure*}

\subsection{Examples of Objects from the Catalog}
\label{sec:eazyexampl}

Figures~\ref{fig:exampleGOODSED} and \ref{fig:exampleBADSED}
provide several examples of photometric data 
fitting by quasar and star templates for
known quasars (for which $\zspec$ is available) or stars.
Figures~\ref{fig:exampleGOODSED}~and~\ref{fig:exampleBADSED} present, respectively, the cases where
one of the quasar spectral templates describes well the
photometric data points and gives a nearly correct estimate 
of $\zphot$ and where quasar template fitting leads
to a misidentification and gives an unreliable estimate of $\zphot$.
The corresponding $\chi^2(z)$. distributions are
shown in the insets.

The objects \textit{3XMM\,J151147.1$+$071406}  ($\zspec =3.481$),
\textit{3XMM\,J001115.2$+$144601} ($\zspec =4.964$), and
 \textit{3XMM\,J004054.6$-$091527}
($\zspec =5.020$) are
examples where the photometric redshift estimation
method works well. In this case, the signal detection
in infrared filters guarantees that the object is not a
star.

The object \textit{3XMM\,J153322.7$+$324351} is an ex-
ample of a poor redshift determination: $\zphot =3.01$.  
The true value measured in the \SDSS\ is $\zspec =1.897$. 

The object \textit{3XMM\,J080630.4+153241} is an M-
type star. It was not included in our catalog of quasar
candidates, because $i^\prime-z^\prime$>0.6 and 
$\chi_{qso}^2/\chi_{star}^2>1$ for
this object. If an object with a similar color relation
is fainter by two apparent magnitudes, i.e., near the
\SDSS\ detection threshold, then the sensitivity of the
\WISE\ and \2MASS\ surveys will be insufficient for its
detection. In that case, the star and quasar templates
will describe the SDSS photometric data points with
comparable $\chi^2$ values, and the star may be mistaken
for a quasar at $\zphot >3$. Nevertheless, many of
such objects can be eliminated using the photometric
constraints (2).

The object \textit{3XMM\,J094109.9+344902} lies at $\zspec = 2.643$. 
There is a \WISE\ source more than
2 arcsec away from the \SDSS\ source. 
Two photometric redshift estimates were made for this object:
by taking into account the fluxes measured in \WISE\ and \2MASS\ filters and using these values only as
upper limits on the infrared flux (see above). In the
former case, $\zphot = 2.62$ (template 12 was used in
fitting) and the object does not enter into the catalog
of quasar candidates, because $\zphot$<2.75. In the
latter case, $\zphot = 3.00$ (template 1 was used) and,
as a result, the object enters into the catalog. Such
cases degrade the catalog purity.

\begin{figure*}[t]
\centering   
\includegraphics[width=0.9\linewidth]{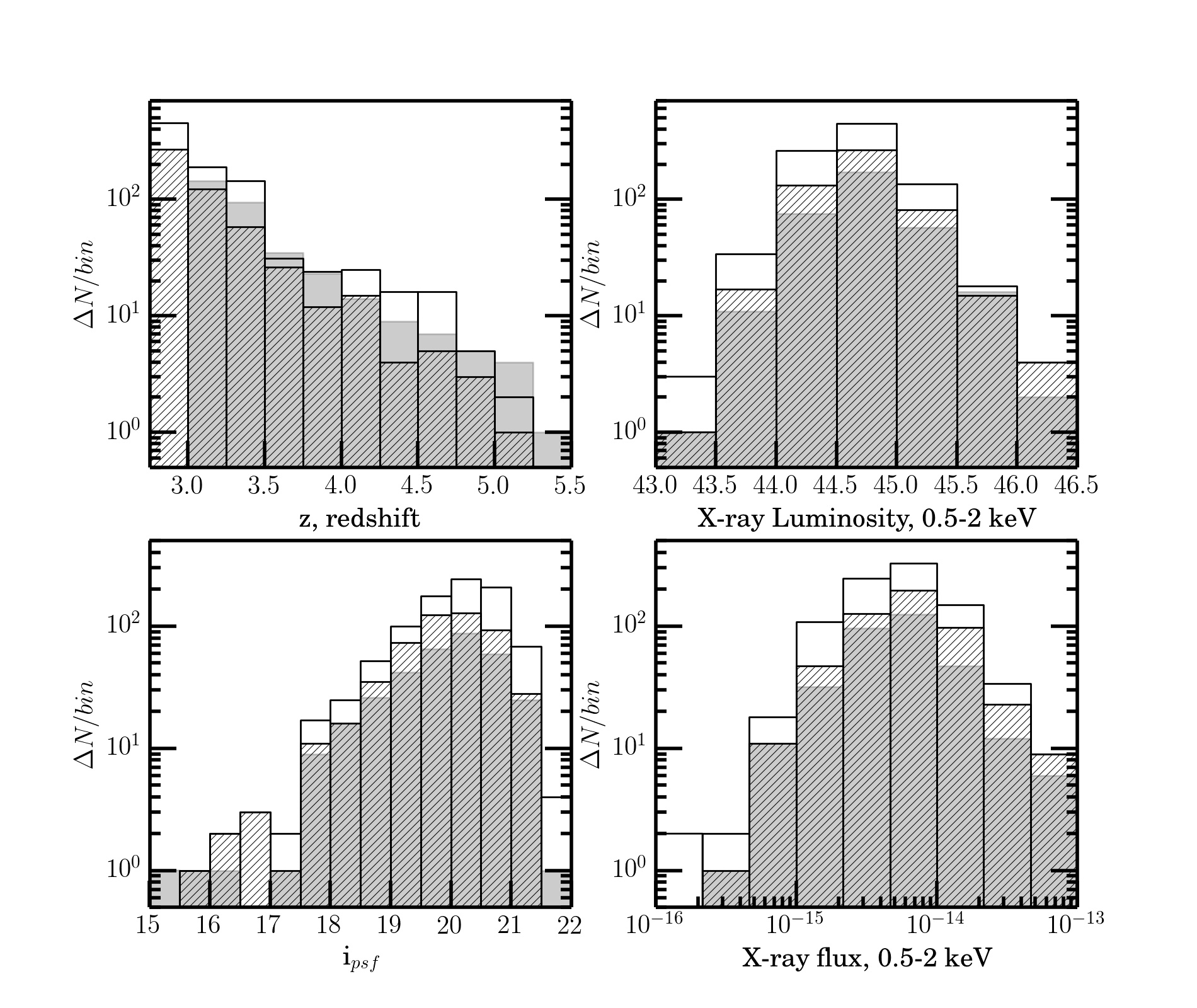}
\caption{Upper left: the distribution of quasar candidates 
  in photometric redshift ($\zphot$). Upper right: the distribution 
  in the common logarithm of X-ray luminosity (in the 0.5--2~keV energy band in the observer’s frame), 
  Lower left: the distribution in $i^\prime $ \SDSS\ magintude. 
  Lower right: the distribution in X-ray 0.5--2~keV flux. 
  The white color indicates the distributions for the entire 
  catalog of quasar candidates. 
  The hatching indicates the objects with $\zphot$ that 
  have known spectroscopic redshifts (with any $\zspec$). 
  The gray color indicates the known quasars with $\zspec >3$. 
  Their distributions were constructed from the spectroscopic redshifts ($\zspec$). 
  The height of each column is the total number of 
  objects of the corresponding subsample in a given bin along the Х axis.}
\label{fig:Catprop} 
\end{figure*}

\section{THE CATALOG AND ITS PROPERTIES}
\label{sec:Catalog}

According to the selection rules described above,
we compiled a catalog of 903 candidates for quasars
at $3<z<5.5$. 
These include 21 3XMM-DR4 pointing targets.
The catalog is presented in the Appendix
in the form of a table where the coordinates, X-ray
fluxes, photometric and spectroscopic (if available)
redshifts, the best quasar and star templates are given. 
The properties of the catalog and its spectroscopic subsample 
are reflected in Fig.~\ref{fig:Catprop}.

The objects in the catalog mostly have photomet-
ric redshifts $2.75\lesssim \zphot \leq3.5$.  
The objects have following distributions by $\zphot$: 
381 at $3\lesssim \zphot \leq4$; 60 at $4<\zphot\leq 5$; 2 at $\zphot >5$. 

The catalog contains 515 known quasars with
reliably measured $\zspec$; 266 of them have $\zspec$>3.
The photometry for these 515 objects is described by
the template of a quasar at $\zphot > 2.75$ and “passes”
according to the selection criteria (see above).
However, 63 known quasars with $\zspec$>3 \citep{flesch15,alam15}
did not satisfy the selection criteria.
Their list is given in a separate table in the Appendix.
Their photometric data points are most often better
described (with smaller $\chi^2$) by the star template 
or the template of a quasar at lower z. Some of the
objects with $\zspec$>3 in the \SDSS\ are classified as
extended sources. No reliable spectroscopic redshift
measurements are available for 388 distant quasar
candidates.
 
The median X-ray flux for the objects from the
catalog (903 sources) is $5.3\times 10^{-15}$~erg\,s$^{-1}$\,cm$^{-2}$
in the 0.5--2~keV energy band. 
All sources with fluxes
above $4\times10^{-14}$~erg\,s$^{-1}$\,cm$^{-2}$
have spectroscopic
redshifts. Because of the constraint on the error in
the z$^\prime$ \SDSS \ magnitude and the requirement
$i^\prime - z^\prime$<0.6, most of our objects turn out to be brighter than
magnitude 21.5 in the $i^\prime$ \SDSS\ filter.

Thus, our catalog of candidates for distant quasars
can (as a result of the subsequent spectroscopic testing) 
lead to a noticeable (up to a factor of 1.5) increase
in the number of known X-ray quasars at $z$>3
relative to the already existing spectroscopic sample (in
the same sky fields).

\section{THE PHOTOMETRIC REDSHIFT ACCURACY,
THE CATALOG COMPLETENESS
AND PURITY}
\label{sec:zspeczphot}
 
The accuracy of the photometric $\zphot$ 
estimates for
distant quasar candidates can be investigated based
on a sample of 329 quasars with known spectro-
scopic redshifts $\zspec$>3 (Fig.~7)
It is convenient
to estimate the scatter of $\zphot$ 
relative to  $\zspec$ (see, e.g., \citealt{hoaglin83},
\citealt{salvato09})
via
the normalized median absolute deviation of
$\Delta z =\left| \zphot - \zspec \right|$: 
 \begin{equation}
 \begin{aligned}
 \sigma_{\Delta z/(1+\zspec)} = %\mbox{ \qquad \qquad \qquad \qquad \qquad \qquad \qquad \qquad} \nonumber          \\ 
  1.48 \times Me \Big (  \frac{\left| \zphot - \zspec \right|}{(1+\zspec)} \Big ),
 %\mbox{NMAD(\zphot \ -- \zspec)} = 1.48 \times Median\left( \left| \mbox{\zphot \ -- \zspec} \right| \right ), 
 \label{eq:nmad}
\end{aligned}
\end{equation}
In our case (for quasars with known $\zspec$>3),
 \SigmaDZ =0.07.  In this case, the percentage
of outliers, the fraction of objects with
$\left| \zphot - \zspec \right|/(1+\zspec)$>0.2 $\approx$ 3\SigmaDZ \ turns out to be $\eta$=\Voutl\%.

Note that the standard confidence intervals for
the EAZY photometric redshift were underestimated
 \citep{dahlen13,hildebrandt08}. 
Only for about of $\approx$30\% the objects from our sample do
the spectroscopic redshifts lie within the 2$\sigma$ error in $\zphot$
(determined in a standard way from the change
in $\chi^2$ near its minimum). Therefore, we will use only
\SigmaDZ\ in discussing the photometric redshift
accuracy.

As regards the completeness of our catalog of
distant quasar candidates, to a first approximation,
it can also be estimated based on the spectroscopic
sample of 329 known quasar at $\zspec$>3.
For this
purpose, consider the ratio of the number of objects
from this sample for which a good photometric redshift 
estimate was obtained, ${|\zphot-\zspec|/(1+\zphot)<0.2}$, to
the total number of objects from the spectroscopic
sample at a given $z$.

The catalog completeness estimated in this way is
about 80\% (the blue solid line in Fig. 8).

\begin{figure}[t]
\centering   
\includegraphics[width=\linewidth]{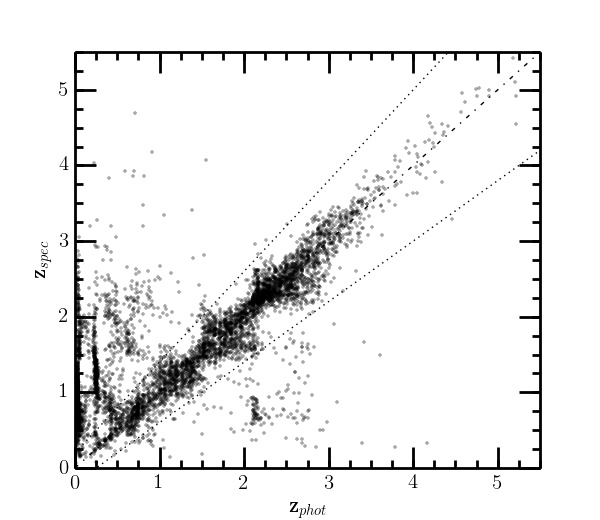}
\caption{Scatter of $\zspec$ relative to $\zphot$ for 6162
  point objects from the spectroscopic \SDSS\ catalog with $\zspec <5.5$
  (having an error $\delta z_{PSF}<0.2$ in the \SDSS\ filter). For
  329 objects with $\zspec >3$ the scatter is \SigmaDZ$=0.07$,
  the percentage of outliers is $\eta=$\Voutl\% (see the determinations in the text). 
  The dotted lines bound the region $|\zphot
  -\zspec|/(1+\zphot)<0.2$.
  The dash--dotted line indicates the straight line $\zphot = \zspec$. }
\label{fig:SDSSzspzph} 
\end{figure}

\begin{figure}[t]
\includegraphics[width=\linewidth]{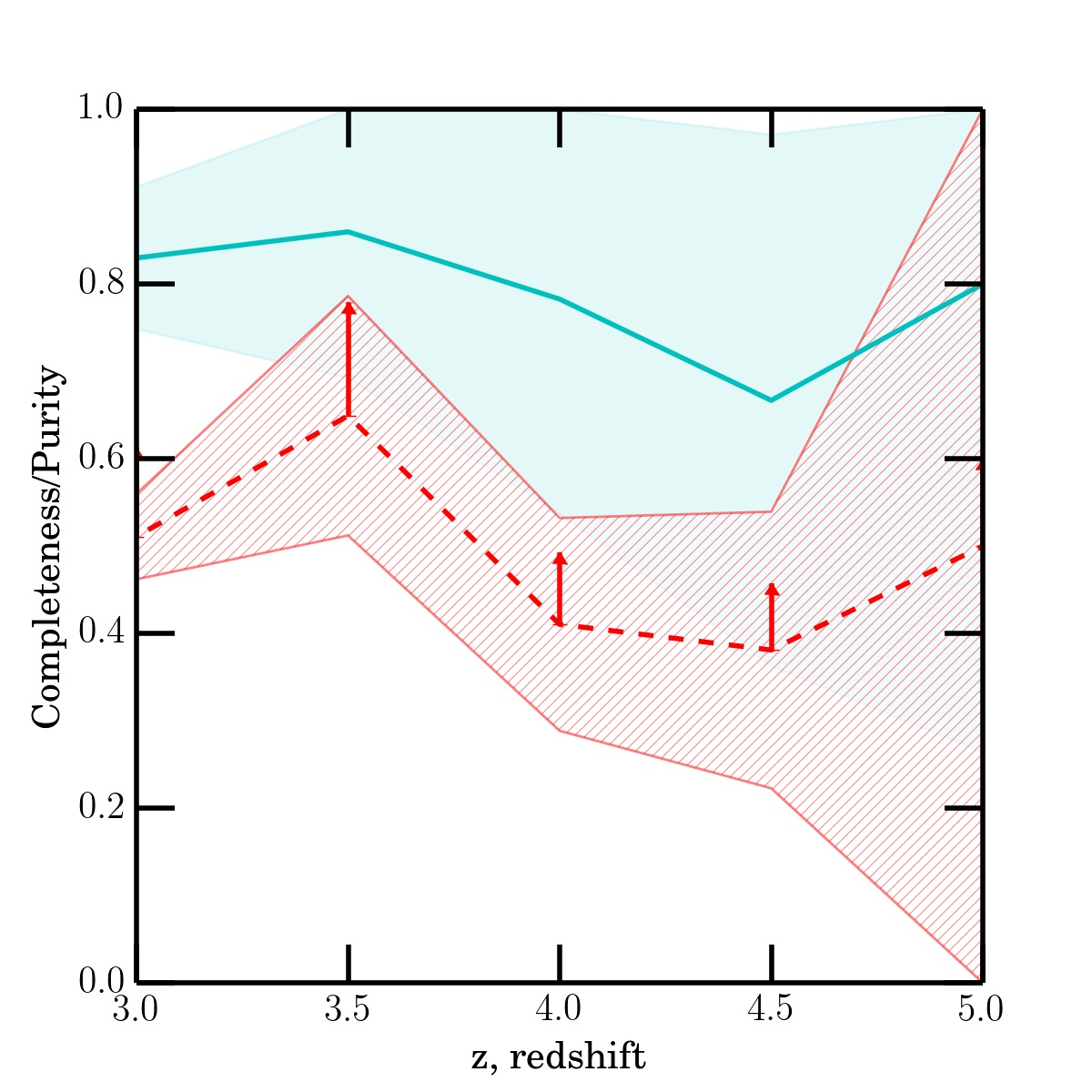}
\caption{Candidate selection completeness 
(the blue solid line and the corresponding $1\sigma$ Poisson uncertainty region) 
and purity
(the red dashed line and the corresponding uncertainty region) 
versus redshift in 0.5-wide bins. 
  The plot was constructed
relative to the spectroscopic sample (329 objects with $3<z<5.5$). 
This dependence was constructed in 0.5-wide redshift
bins. The arrows mean that the purity relative
 to the spectroscopic sample 
 should be considered as a lower limit of the true
purity, which is not yet known.
}
\label{fig:SDSScomplreliab}
\end{figure}

However, it is clear that this estimate cannot
be deemed quite reliable, because the spectroscopic
sample used was compiled only from optical data and
can differ significantly in its properties from the X-ray
quasar sample. Indeed, in our catalog, which was
compiled on the basis of an X-ray survey, there are
much more objects per unit sky area, some of which
can be quasars at $z$> 3. Moreover, in estimating the
completeness with respect to all X-ray quasars (at a
given X-ray flux), we should also take into account
the objects that are seen as X-ray sources in the
3XMM-DR4 survey but are too faint in the optical
band and, therefore, enter neither into our catalog
nor into the spectroscopic sample. To take into
account all these possibilities requires performing the
corresponding simulations of the quasar population,
which we are going to do later on.

It will be possible to investigate the catalog purity,
i.e., the ratio of the number of true quasars at $z>3$
to the number of all objects in the catalog, later on
by performing spectroscopic observations of a 
representative subsample of objects from our catalog.
By considering the ratio of the number of quasars
from the spectroscopic sample that entered into our
catalog to the number of all objects in the catalog, we
can obtain a lower limit on the purity of our catalog,
which is indicated in Fig.~\ref{fig:SDSScomplreliab} by the red dashed line.

\section{THE X-RAY SURVEY AREA AND THE $L\lowercase{og} N$--$L\lowercase{og} S$ CURVE}
\label{sec:vyborkaarea}

\begin{figure}
  \centering
  \includegraphics[width=\linewidth]{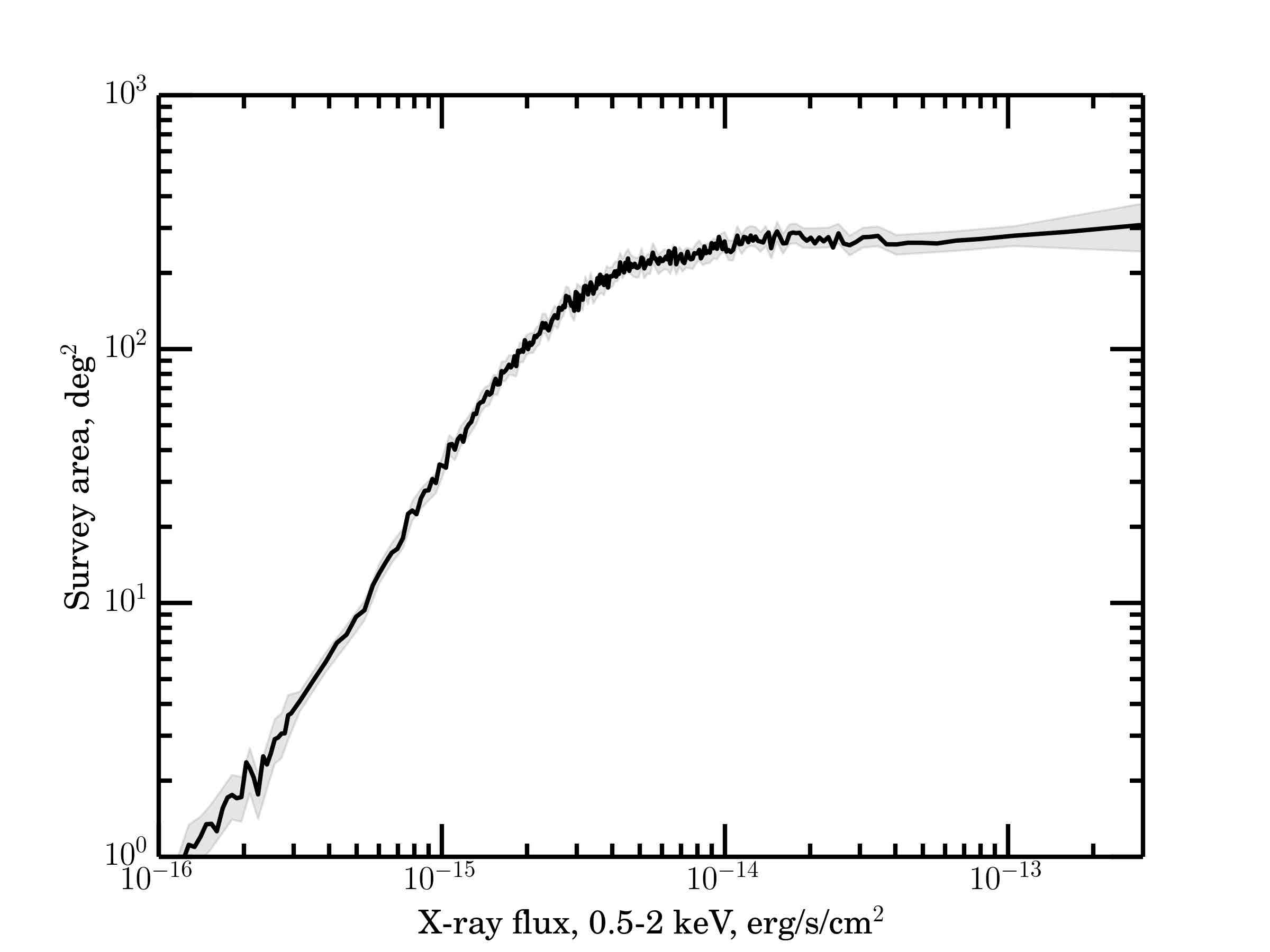}
  \caption{Survey area (\3XMMDR4) versus X-ray flux (0.5-2~кэВ) 
  in regions with \SDSS\ coverage at Galactic latitudes $|b|>20^\circ$. 
  The gray region indicates the $2\sigma$ Poisson errors.}
  \label{fig:Areasurvey}
\end{figure}

The total geometric area of the overlap between
the
\3XMMDR4\ X-ray survey and the photometric \SDSS\ is about $\approx$300~deg$^2$
at Galactic latitudes $|b|>20^\circ$. 
The survey area as a function of the X-ray flux
can be estimated as the ratio of the total number of
\3XMMDR4\ X-ray sources in \SDSS\ 
fields with an
X-ray flux above a given one to the expected density
of such sources in the sky. The expected density
of sources at a given flux was calculated using the
approximation from
\cite{mateos08}, where the
statistics of \XMM\ sources on a total area
of 132~deg$^2$
at high Galactic latitudes, $|b|>20^\circ$, was
analyzed in detail. The area estimate obtained in this
way is presented in Fig.~\ref{fig:Areasurvey}.

Using this estimate of the survey area as a function
of the X-ray flux, we can construct the $\log N$--$\log S$
curve for quasar candidates at $z>3$.
To take into
account the fact that not all quasars at $z>3$
 among the \3XMMDR4\
X-ray sources are also seen in the
optical \SDSS\ images, the $\log N$--$\log S$ curve can
be properly corrected. For this purpose, we can use
the estimate of the fraction of X-ray quasars at $z>3$
having \SDSS\ photometry at a given X-ray flux that
was discussed above (see Fig.~\ref{fig:DR12fluxtotal}).
The $\log N$--$\log S$ curve for quasars at $z>3$
was constructed as follows:
\begin{equation}
\mbox{$N(>S)$} =\sum_{i=1}^{N_{S}} \frac{1}{\Omega_i \alpha_i} \ ,
\label{eq:logNlogSour}
\end{equation}
with the corresponding error being
\begin{equation}
  \sigma_N = \sqrt{ \sum_{i=1}^{N_{S}} \left[ \frac{1}{\Omega _i^2 \alpha _i^2 }  + \left(\frac{\delta \alpha _i}{\Omega _i \alpha _i^2 }\right)^2 \right] } \ ,
\label{eq:logNlogSourerr}
\end{equation}
where $\Omega _i \equiv \Omega(S_i)$ is the survey area at flux $S_i$,
$\alpha_i\equiv\alpha(S_i)$ is the fraction of X-ray quasars at $z>3$,
that have \SDSS\ photometry, and the summation is over
$N_{S}$-sources with fluxes $S_i \geq S$.

\begin{figure}
\includegraphics[width=\linewidth]{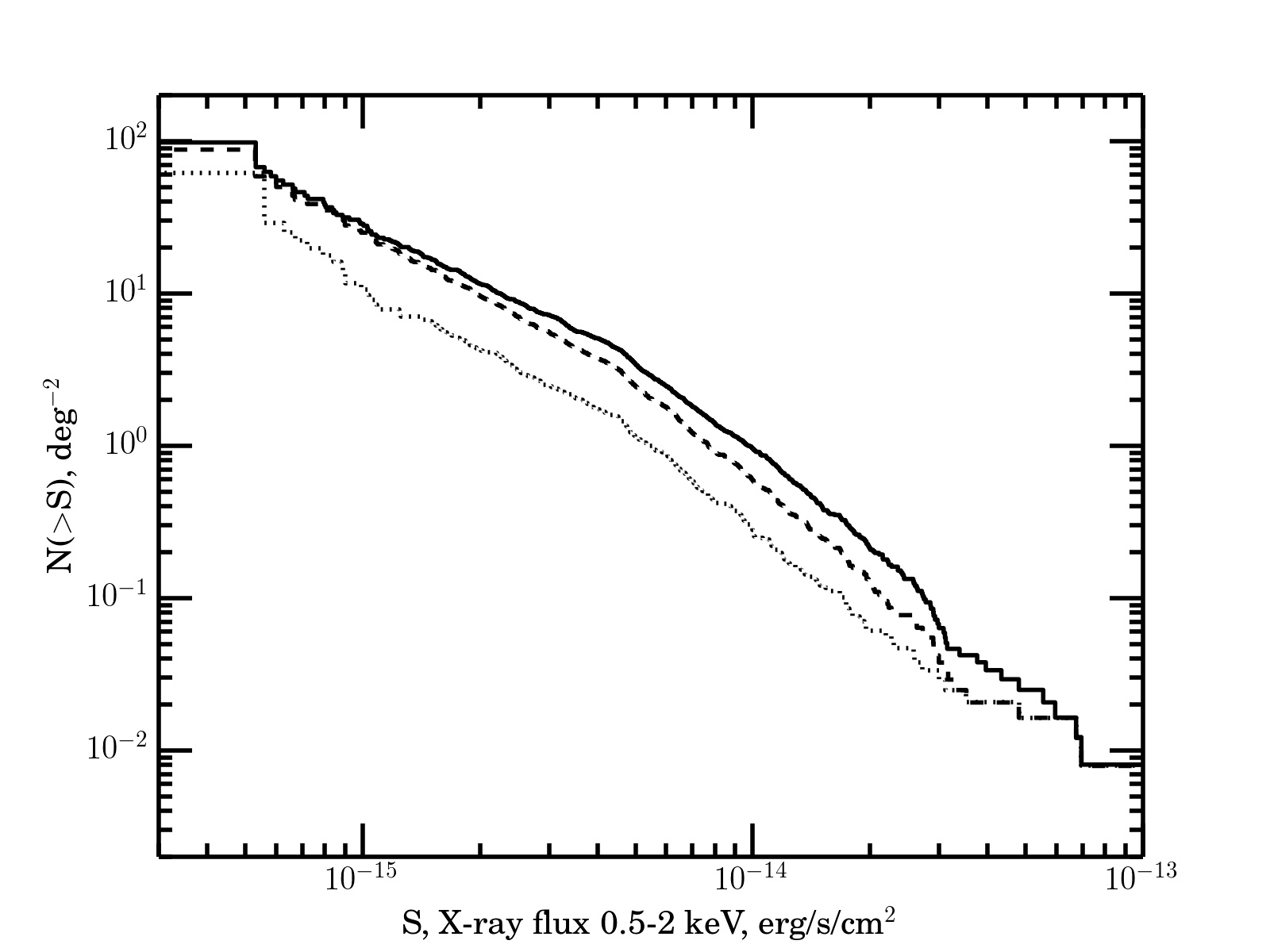}
\caption{The black solid line indicates the $Log N$--$Log S$
  distribution of candidates for quasars at $z>3$.  
  (i.e., with $\zphot$>2.75).
  The dotted and dashed lines indicate, respectively, the
  $Log N$--$Log S$ distribution only for the objects from this sample with
   known spectroscopic redshifts $\zspec >3$ and for the objects with
   $\zspec >3$ or without $\zspec$.
   The true $Log N$--$Log S$ curve for
   quasars at $z>3$ must lie between these two curves.}
\label{fig:ResLogNlogS}
\end{figure}

The $\log N$--$\log S$ curve constructed from all 
objects in our catalog is indicated in Fig.~\ref{fig:ResLogNlogS}
by the black solid line. However, this curve actually does
not reflect the true distribution of quasars at $z>3$ 
in X-ray flux, because quite a few objects with redshifts
entered into our catalog and \Nexclu\ quasars with $\zspec >3$.  
were missed.
Therefore, Fig.~\ref{fig:ResLogNlogS} shows two
more $\log N$--$\log S$ curves: only for the sources with $\zspec >3$ 
(dotted lines) and for the sources that do not
have either the spectrum or $\zspec >3$ (dashed line).
The true $\log N$--$\log S$ curve must lie between these
two curves.

\section{CONCLUSIONS}

We searched for quasar candidates at $z>3$ in the
\3XMMDR4\ catalog of X-ray sources using a photometry of
 \SDSS, \2MASS, \WISE. A catalog of 903 candidates for 
 quasars with photometric redshifts $\zphot$>2.75 was 
 compiled.

Note that several catalogs of candidates for quasars
with photometric redshift estimates have already been
produced previously based on data from the photometric 
\SDSS\ catalog (see, e.g., \cite{bovy12,richards09}),
and WISE data have also been
used in their recent paper by \cite{dipompeo15}.
Our quasar selection method differs fundamentally in
that the selection results
(the density of quasars in the
sky) can be presented in the form of a dependence on
the X-ray flux, which is important for the subsequent
construction of the X-ray luminosity function. A
preliminary comparison showed that our sample of
candidates for quasars at $z\sim3$ is essentially identical
(in the sky fields we investigated) to the sample
from \cite{dipompeo15}. 
However, there are
significant discrepancies between the samples at $z\gtrsim$4 
whose degree and cause are to be investigated
in a subsequent work.

Preliminary estimates of the selection quality
using the \SDSS\ DR12 catalog of spectroscopic
redshifts (supplemented by quasars from the catalog
published recently by \cite{flesch15}) для рентгеновских
источников \XMM\ X-ray sources (\3XMMDR4) showed that
the completeness of our sample is about 80\% and
its purity is at least 50\%. 
However, these estimates
are not quite reliable and must be checked in future
through simulations of the quasar population and using 
optical observations of objects from this sample.

The accuracy of the photometric redshift estimates
is \SigmaDZ =0.07 with $\approx$\Voutl\% of outliers. 
This
accuracy is comparable to other similar measurements 
based on photometric \SDSS\ data \citep[see, e.g.,][]{kitsionas05}. 
Such an accuracy is sufficient
to measure the luminosity function. We are going to
perform these measurements in our next work.

About half of the bright objects in our sample
have measured redshifts. The objects in our 
sample have apparent magnitudes $i^\prime\lesssim 21.5$.
We began the spectroscopic observations of the remaining
objects from this sample with the 6-m telescope at
the Special Astrophysical Observatory of the Russian
Academy of Sciences and with the 1.6-m telescope
at the Sayan Observatory of the Institute of Solar–
Terrestrial Physics, the Siberian Branch of the Russian 
Academy of Sciences, where a new medium- and
low-resolution spectrograph well suited for such observations 
was installed \citep{burenin16}
These
observations not only will allow the sample quality
to be estimated but also will probably allow a large
number of quasars at $z>$3 to be confirmed, which, in
turn, will make it possible to measure more accurately
the X-ray luminosity function for quasars.

It will be possible to apply the high-redshift quasar
selection algorithms tested in our paper to select such
objects among the X-ray sources of the SRG all-sky
survey. As can be seen, in particular, from the results
of our paper, it will be possible to detect of the order of
several tens of thousands of quasars at 3 < z < 6 in
the SRG survey. This unique observational material
will allow the growth history of supermassive black
holes at the epoch of active galaxy formation to be
studied in detail.
 
\paragraph{ACKNOWLEDGMENTS}  
This study was supported by the Russian Science
Foundation (project no. 14-22-00271).
We would
like to thank V. Astakhov for translation of the
paper in English.
The electronic version of the article's tables is available \footnote{http://adsabs.harvard.edu/abs/2016yCat..90420313K} at 
the VizieR catalogue access tool, CDS,
Strasbourg, France \citep{ochsenbein00}.

\vfill
\eject
 
%\pagebreak

%%% Список литературы

\onecolumn

\begin{center}

%\begin{landscape}
\section*{Appendix}
\subsection*{Catalog of candidates for quasars at 3<z<5.5. The table of redshifts.}  
  { \tiny \setlength{\tabcolsep}{3pt}
  \smallskip 
  \begin{longtable}{lccrrrrclcccccccccccccc} 
  \hline
	\hline
	\multicolumn{1}{l}{N} &
	\multicolumn{1}{c}{NAME 3XMM} &
	\multicolumn{1}{c}{SRCID} &
	\multicolumn{1}{c}{RA} &
	\multicolumn{1}{c}{DEC} &
	\multicolumn{1}{r}{F$_{0.5-2}^{-14}$} &
	\multicolumn{1}{r}{$\Delta$F$_{0.5-2}^{-14}$} &
	\multicolumn{1}{l}{$i^\prime_{PSF}$} &
	\multicolumn{1}{l}{$z^\prime_{PSF}$} &
	\multicolumn{1}{l}{$z_{\mbox{spec}}$} &
	\multicolumn{1}{l}{$z_{ref}$} &
	\multicolumn{1}{l}{$z_{\mbox{phot}}$} &
	\multicolumn{1}{l}{FL} &
	\multicolumn{1}{l}{T$_{Q}$} &
	\multicolumn{1}{l}{T$_{S}$} &
	\multicolumn{1}{l}{NB} 
	\\
	
	\multicolumn{1}{l}{(1)} &
	\multicolumn{1}{c}{(2)} &
	\multicolumn{1}{c}{(3)} &
	\multicolumn{1}{c}{(4)} &
	\multicolumn{1}{c}{(5)} &
	\multicolumn{1}{c}{(6)} &
	\multicolumn{1}{c}{(7)} &
	\multicolumn{1}{l}{(8)} &
	\multicolumn{1}{l}{(9)} &
	\multicolumn{1}{l}{(10)} &
	\multicolumn{1}{l}{(11)} &
	\multicolumn{1}{l}{(12)} &
	\multicolumn{1}{l}{(13)} &
	\multicolumn{1}{c}{(14)} &
	\multicolumn{1}{c}{(15)} &
	\multicolumn{1}{l}{(16)} 
	\\
	\hline
	\endhead
	
	\hline \\
	\endfoot
	
	\blfootnote{NAME is the name in the 3XMM-DR4 (3XMMJ...), SRCID is the unique X-ray source number in the 3XMM-DR4, RA is the right ascension and DEC is the declination in degrees in the 3XMM-DR4,  F$_{0.5-2}^{-14}$ and $\Delta$F$_{0.5-2}^{-14}$ are the 0.5--2~keV flux and its error ($\times 10^{-14}$ erg/s/cm$^2$), $i^\prime_{PSF}$ is the apparent magnitude in the $i^\prime$ SDSS (AB, PSF), $z^\prime_{PSF}$ is the apparent
	magnitude in the $z^\prime$ SDSS filter (AB, PSF),   
	$z_{\mbox{spec}}$ is the spectroscopic redshift, $z_{ref}$ is the number of the reference to the paper with
published  $z_{\mbox{spec}}$ (0 --- SDSS DR12, the zWarning flag is given in parentheses; 1 --- the catalog of X-ray sources by Flesch (2015)the
number of the data source from Flesch (2015) is given in parentheses), $z_{\mbox{phot}}$ is the photometric redshift, FL is the data flag (D --- there
is another optical source within the 2$\sigma$ position error of the XMM X-ray source;
 w --- the $z_{\mbox{phot}}$ estimates were obtained only from
the SDSS photometry, although there is a WISE source nearby; t --- the XMM pointing target XMM, r --- the star template describes the
photometry better than does the quasar template ($\chi^2_{Star}/\chi^2_{QSO}<1$), T$_{Q}$ is the
template number from the quasar library, T$_{S}$ is the star template (Pickles 1998) with the smallest $\chi^2$ , NB --- число фотометрических полос, используемых при аппроксимации. NB is the number of photometric
bands used in fitting.

	}%\footnote{http://adsabs.harvard.edu/abs/2016yCat..90420313K}
	
	\\
1&J000335.5$-$061112&119255&0.8981&$-$6.1869&2.851&0.205&19.25&19.00&2.8569&0(0)&3.05&&4&f8i&7\\
2&J000303.7+020416&119783&0.7655&2.0714&1.295&0.211&20.76&20.33&2.9469&0(0)&3.02&&12&g0iii&8\\
3&J000444.2+020801&117610&1.1846&2.1337&0.573&0.229&20.95&20.69&2.6655&0(0)&2.80&&1&a7v&5\\
4&J000531.3+000840&117756&1.3806&0.1445&0.963&0.281&20.99&20.58&2.8480&0(0)&2.80&&7&f5iii&7\\
5&J000443.6$-$084036&117599&1.1820&$-$8.6767&0.944&0.323&20.30&20.12&&&3.85&&3&k3i&5\\
6&J000511.6$-$084201&117868&1.2986&$-$8.7005&0.498&0.185&19.64&19.64&3.2059&0(0)&3.30&&1&f8i&7\\
7&J000533.7$-$084825&117722&1.3908&$-$8.8071&0.577&0.232&19.24&19.46&&&2.82&&7&f8iv&8\\
8&J000618.1$-$084410&117240&1.5757&$-$8.7362&0.872&0.433&20.45&20.57&3.3234&0(0)&3.15&&7&g0i&7\\
9&J000942.5+125251&118363&2.4272&12.8811&3.149&0.528&20.21&20.03&2.3926&0(0)&2.82&&3&g8v&7\\
10&J001033.5+105231&118456&2.6398&10.8755&1.171&0.416&20.91&20.48&2.9202&0(0)&2.93&&3&rk0v&7\\
11&J001037.5+105526&118489&2.6565&10.9239&1.463&0.406&20.25&20.04&2.6765&0(0)&2.81&&1&f2v&5\\
12&J001115.2+144601&117909&2.8135&14.7671&10.810&0.299&18.28&18.10&4.9643&0(0)&4.54&t&5&m2i&9\\
13&J001340.4+054554&121657&3.4183&5.7651&0.927&0.213&19.95&19.85&2.7123&0(0)&2.97&&1&f2iii&7\\
14&J001504.7+171517&121325&3.7700&17.2549&0.160&0.069&20.63&20.70&3.1627&0(0)&2.84&&7&f6v&5\\
15&J001756.7+163007&122306&4.4864&16.5021&0.183&0.070&20.29&20.29&3.5487&0(0)&3.83&&1&k3iii&6\\
16&J001920.9+163209&122068&4.8372&16.5359&0.859&0.167&19.04&19.00&&&2.96&&1&f6v&8\\
17&J001754.1+161406&122310&4.4755&16.2350&0.465&0.136&20.81&20.45&&&2.87&&3&wg5iii&8\\
18&J002015.9+214914&121995&5.0664&21.8208&0.346&0.120&19.72&19.51&&&3.03&&12&k3iii&9\\
19&J002159.9$-$083503&120127&5.4996&$-$8.5843&0.236&0.092&20.78&20.72&&&2.84&&12&rf6v&9\\
20&J002127.3$-$020333&120370&5.3640&$-$2.0593&1.829&0.154&17.00&16.90&2.596&1(824)&2.77&t&1&wf5v&12\\
21&J002150.6$-$150427&119878&5.4611&$-$15.0742&0.189&0.047&21.31&20.72&&&2.79&&7&f5iii&5\\
22&J002208.0$-$150540&120066&5.5334&$-$15.0945&1.635&0.103&18.80&18.58&4.52&1(401)&4.40&t&4&m2i&7\\
23&J002244.4+013250&121123&5.6854&1.5474&0.915&0.105&19.51&19.38&2.9823&0(0)&2.76&&2&f8iv&7\\
24&J002527.5+105118&120782&6.3648&10.8553&0.337&0.157&19.63&19.33&&&3.13&&3&rk0iii&8\\
25&J002630.2+165656&113930&6.6259&16.9490&1.510&0.196&20.23&19.97&2.8593&0(0)&2.86&&4&f5i&7\\
26&J002726.4+170730&113471&6.8602&17.1251&0.377&0.091&20.05&20.17&3.8857&0(0)&4.02&&1&k2i&5\\
27&J002706.9+261559&113871&6.7791&26.2666&0.943&0.376&20.32&19.82&&&3.29&&12&rk4iii&8\\
28&J003027.7+261355&19579&7.6157&26.2324&0.239&0.044&19.78&19.93&3.2136&0(0)&3.11&&7&g0i&8\\
29&J003057.9+261744&16416&7.7415&26.2956&0.900&0.076&20.45&20.02&&&3.01&&3&wg5iii&7\\
\\
887&J233329.9+152539&237875&353.3750&15.4276&0.426&0.155&19.00&18.84&3.6508&0(0)&3.56&&5&rk3iii&8\\
888&J233633.9+210442&238644&354.1413&21.0784&0.220&0.063&21.05&20.50&&&2.98&&3&g8v&8\\
889&J233641.2+211955&238680&354.1717&21.3322&0.365&0.095&20.17&20.14&&&2.92&&3&g8v&8\\
890&J234214.1+303606&237645&355.5590&30.6017&0.799&0.203&19.92&19.79&&&3.20&wD&2&wk2iii&5\\
891&J234309.6+001325&237594&355.7901&0.2239&0.470&0.110&20.59&20.83&2.8240&0(0)&2.96&&9&f5i&5\\
892&J234448.7$-$042651&241540&356.2031&$-$4.4475&0.197&0.132&18.59&18.44&&&2.88&D&3&rg5iii&9\\
893&J234455.3+092444&241510&356.2304&9.4125&0.163&0.143&20.79&20.85&&&3.18&D&12&rg5iii&7\\
894&J234752.5+010306&242170&356.9688&1.0518&0.446&0.164&20.62&20.46&2.4114&0(0)&2.86&&12&g0v&8\\
895&J234916.0+020722&241813&357.3167&2.1230&0.732&0.208&20.87&20.59&2.6551&0(0)&2.77&&4&f0iii&5\\
896&J234956.8+363237&239836&357.4871&36.5437&0.548&0.203&19.55&19.40&&&2.92&&3&wg8iii&8\\
897&J235054.6+200939&239972&357.7276&20.1609&1.000&0.121&18.49&18.47&3.1629&0(0)&2.88&&6&f8i&12\\
898&J235201.3+200901&239304&358.0055&20.1506&1.163&0.151&18.37&18.33&3.0794&0(0)&3.09&&1&g0iii&11\\
899&J235101.0+203504&240044&357.7544&20.5847&0.530&0.137&19.99&19.95&&&2.80&&5&g0iv&8\\
900&J235111.2+202016&239889&357.7970&20.3378&0.996&0.358&20.11&19.86&2.329&0(0)&2.76&&1&f02iv&7\\
901&J235435.5$-$101513&239560&358.6481&$-$10.2537&1.159&0.232&19.30&19.19&3.1203&0(0)&2.97&&1&f5v&7\\
902&J235502.9+060825&240542&358.7623&6.1405&1.497&0.408&19.04&18.80&2.7077&0(0)&2.79&&7&f8i&8\\
903&J235555.9+055933&240529&358.9831&5.9926&0.344&0.121&20.22&20.14&&&3.35&&2&wk2iii&8\\
	\hline
	\hline
\end{longtable}
}
\newpage

\subsection*{Known quasars with $\zspec$ >3 from the \3XMMDR4 catalog that did not enter into our catalog of candidates}
{
\tiny
\setlength{\tabcolsep}{3pt}
\smallskip
\begin{longtable}{lccrrrrclcccccccccccccc}
	\hline
	\hline
	\multicolumn{1}{l}{N} &
	\multicolumn{1}{c}{NAME 3XMM} &
	\multicolumn{1}{c}{SRCID} &
	\multicolumn{1}{c}{RA} &
	\multicolumn{1}{c}{DEC} &
	\multicolumn{1}{r}{F$_{0.5-2}^{-14}$} &
	\multicolumn{1}{r}{$\Delta$F$_{0.5-2}^{-14}$} &
	\multicolumn{1}{l}{$i^\prime_{PSF}$} &
	\multicolumn{1}{l}{$z^\prime_{PSF}$} &
	\multicolumn{1}{l}{$z_{\mbox{spec}}$} &
	\multicolumn{1}{l}{$z_{ref}$} &
	\multicolumn{1}{l}{$z_{\mbox{phot}}$} &
	\multicolumn{1}{l}{FL} &
	\multicolumn{1}{l}{T$_{Q}$} &
	\multicolumn{1}{l}{T$_{S}$} &
	\multicolumn{1}{l}{NB} 
	\\
	
	\multicolumn{1}{l}{(1)} &
	\multicolumn{1}{c}{(2)} &
	\multicolumn{1}{c}{(3)} &
	\multicolumn{1}{c}{(4)} &
	\multicolumn{1}{c}{(5)} &
	\multicolumn{1}{c}{(6)} &
	\multicolumn{1}{c}{(7)} &
	\multicolumn{1}{l}{(8)} &
	\multicolumn{1}{l}{(9)} &
	\multicolumn{1}{l}{(10)} &
	\multicolumn{1}{l}{(11)} &
	\multicolumn{1}{l}{(12)} &
	\multicolumn{1}{l}{(13)} &
	\multicolumn{1}{c}{(14)} &
	\multicolumn{1}{c}{(15)} &
	\multicolumn{1}{l}{(16)} 
	\\
	%\multicolumn{1}{l}{(22)} \\
	%\multicolumn{1}{l}{(23)} &
	%\multicolumn{1}{l}{(24)} \\
	%\multicolumn{1}{l}{(25)} &
	%\multicolumn{1}{l}{(26)} \\
	\hline
	\endhead
	
	\hline \\
	\endfoot
	
1&J001049.0+290139&118405&2.7046&29.0277&0.275&0.192&20.23&20.18&3.370&0(0)&3.18&r&5&g2i&5\\
2&J002654.9+171944&113822&6.7289&17.3289&1.125&0.132&21.21&20.79&3.095&0(0)&2.72&&12&f8i&7\\
3&J005156.6+272847&128940&12.9862&27.4799&0.289&0.183&19.73&19.57&3.331&0(0)&3.46&r&1&g5iv&7\\
4&J011117.7+325832&44045&17.8243&32.9758&0.245&0.099&19.59&19.39&3.205&1(1411)&0.69&&13&m5iii&9\\
5&J011544.8+001513&125421&18.9371&0.2538&0.194&0.064&21.53&20.96&5.1&1(1245)&0.61&&14&m4v&7\\
6&J020231.1$-$042246&43456&30.6301&$-$4.3794&1.285&0.279&20.68&20.43&4.270&0(0)&0.94&G&14&m7iii&9\\
7&J020253.7$-$065043&92295&30.7240&$-$6.8453&0.155&0.083&20.69&20.34&3.866&0(0)&0.79&G&13&m5iii&8\\
8&J020702.2$-$065233&96071&31.7593&$-$6.8759&0.366&0.111&20.51&20.45&3.065&0(0)&2.66&&6&f0iii&7\\
9&J021338.6$-$051615&20962&33.4110&$-$5.2712&0.929&0.096&20.88&20.22&4.544&0(0)&0.55&&14&m4iii&7\\
10&J021343.2$-$042042&107259&33.4301&$-$4.3450&0.347&0.163&20.45&20.38&3.074&0(0)&0.06&&7&m5iii&7\\
11&J021504.1$-$040704&107902&33.7671&$-$4.1178&0.636&0.208&20.53&20.34&3.283&0(0)&0.03&&6&f8i&7\\
12&J022048.5$-$033711&110418&35.2025&$-$3.6197&0.564&0.180&21.42&20.61&3.116&0(0)&2.71&&12&f8i&7\\
13&J022251.7$-$050713&103134&35.7157&$-$5.1203&0.880&0.122&20.01&19.57&3.86&1(1758)&0.67&&13&m3ii&8\\
14&J022645.4$-$043616&102470&36.6893&$-$4.6046&0.726&0.152&18.45&17.89&3.295&1(559)&0.49&&14&m3ii&12\\
15&J022706.4$-$041924&105904&36.7769&$-$4.3235&0.392&0.100&20.65&20.60&3.285&1(650)&2.90&r&3&rg0v&5\\
16&J022849.6$-$043946&106357&37.2069&$-$4.6630&0.241&0.187&21.04&20.60&3.211&0(0)&2.62&&7&f5i&7\\
17&J023226.0$-$053729&160738&38.1086&$-$5.6248&0.532&0.258&18.83&18.60&4.564&0(0)&0.28&r&14&m2iii&8\\
18&J024923.6$-$040206&163652&42.3484&$-$4.0351&0.391&0.195&20.81&20.62&3.029&0(0)&2.94&r&3&g5v&5\\
19&J030222.0+000630&155293&45.5921&0.1085&0.708&0.072&20.65&20.50&3.306&0(0)&0.03&&13&k0iii&8\\
20&J030435.4$-$000250&156232&46.1476&$-$0.0474&0.605&0.241&20.41&20.16&3.055&0(0)&0.03&&13&f8i&8\\
21&J080334.9+391923&79784&120.8954&39.3232&0.348&0.151&20.37&20.29&3.011&0(0)&2.67&&7&f5i&7\\
22&J085026.7+630020&307591&132.6113&63.0057&0.647&0.195&18.36&18.23&3.363&0(0)&3.08&r&1&k3iii&11\\
23&J085809.3+274228&300636&134.5391&27.7080&0.246&0.052&20.39&20.48&3.464&0(0)&3.10&r&3&g8iv&5\\
24&J093521.2+612339&317819&143.8387&61.3943&0.693&0.101&20.04&20.12&4.042&0(0)&0.30&&14&m2iii&7\\
25&J094013.9+344628&321396&145.0579&34.7747&2.295&0.257&21.35&20.90&3.355&0(0)&0.38&&13&rk4iii&7\\
26&J094404.1+165056&322206&146.0174&16.8489&0.516&0.147&20.27&20.28&3.017&0(0)&3.05&r&1&f5iii&5\\
27&J095752.0+015119&313069&149.4669&1.8555&0.254&0.107&21.02&20.40&4.174&1(1886)&0.90&DG&14&m7iii&7\\
28&J100055.3+250907&310877&150.2306&25.1521&0.124&0.040&21.17&20.49&3.2&1(1538)&0.41&G&13&m3iii&8\\
29&J100226.1+024611&23727&150.6090&2.7687&0.428&0.057&20.06&20.19&3.038&1(271)&3.13&Gr&6&g2iv&5\\
30&J100655.8+050325&315437&151.7327&5.0571&1.454&0.644&19.46&19.39&3.086&0(0)&3.60&r&1&rk0v&8\\
31&J104445.6$-$011756&283334&161.1901&$-$1.2990&0.248&0.071&20.35&20.29&3.496&0(0)&3.73&r&1&g5i&5\\
32&J104920.9+510041&16951&162.3372&51.0114&0.171&0.034&20.87&20.77&3.057&0(0)&0.04&&7&f5i&7\\
33&J104808.3+583718&275930&162.0347&58.6217&1.136&0.348&20.31&20.13&3.285&0(0)&0.05&&13&wk2iii&8\\
34&J111151.3+061321&44028&167.9642&6.2227&0.146&0.031&20.18&20.33&3.200&0(0)&3.13&r&6&g0iv&5\\
35&J115659.3+551309&27772&179.2477&55.2190&0.933&0.234&18.39&18.09&3.110&0(0)&0.23&&13&m1iii&12\\
36&J120613.4+443527&355201&181.5561&44.5908&0.564&0.114&20.93&20.28&3.483&0(0)&0.80&&13&m5v&9\\
37&J120735.5+251140&356435&181.8982&25.1947&0.538&0.165&20.21&20.14&3.008&0(0)&2.73&&7&f5i&7\\
38&J122652.2+013632&349269&186.7178&1.6090&0.687&0.201&19.61&19.74&3.015&0(0)&2.70&&6&f02iv&8\\
39&J123142.1+110308&23419&187.9259&11.0522&0.382&0.045&20.06&20.00&3.024&0(0)&2.60&&5&f02iv&7\\
40&J123231.0+121846&348272&188.1293&12.3131&0.302&0.153&20.57&20.69&3.055&0(0)&2.75&&7&f5iv&7\\
41&J123613.4+275151&350816&189.0561&27.8643&0.424&0.106&20.10&20.07&3.780&0(0)&0.70&&13&m3ii&11\\
42&J123752.6+092934&350323&189.4692&9.4930&0.925&0.216&19.60&19.54&3.027&0(0)&2.67&&7&f02iv&7\\
43&J124405.1+125757&368647&191.0214&12.9659&1.101&0.200&21.15&20.86&3.1&1(611)&0.43&D&13&m2iii&7\\
44&J124825.2+673135&366095&192.1052&67.5266&0.519&0.101&19.15&19.10&3.224&0(0)&2.73&r&9&f8i&5\\
45&J130200.0+281213&362403&195.5002&28.2038&0.163&0.080&20.74&20.45&3.084&0(0)&2.68&&2&f5iii&8\\
46&J130206.5+281117&362291&195.5272&28.1882&0.249&0.098&21.16&20.34&4.884&0(0)&5.55&&12&m5v&8\\
47&J131047.8+322518&364785&197.6995&32.4219&0.090&0.048&20.34&19.96&3.009&0(0)&0.70&&8&f0i&9\\
48&J132451.8+032722&331382&201.2159&3.4563&0.641&0.172&20.81&20.19&3.019&0(0)&3.22&&12&wk2iii&7\\
49&J133002.2+241634&53129&202.5096&24.2766&0.192&0.055&20.75&20.21&3.038&0(0)&2.73&&12&f2ii&7\\
50&J135613.7+182358&328590&209.0575&18.3996&0.482&0.087&21.10&20.68&3.927&0(0)&0.68&G&13&m3ii&8\\
51&J140149.8+024835&11017&210.4580&2.8100&0.982&0.073&21.19&20.53&3.83&1(643)&0.26&G&7&m2iii&7\\
52&J142437.8+225601&346440&216.1579&22.9338&41.066&10.834&15.39&15.34&3.62&1(1417)&3.57&G&7&rk3iii&12\\
53&J143023.7+420436&15580&217.5989&42.0768&54.979&0.640&19.16&19.31&4.656&0(0)&4.15&tr&1&m4v&5\\
54&J150603.5+012757&41624&226.5147&1.4664&0.787&0.136&20.78&20.36&3.852&0(0)&3.41&G&12&m2iii&7\\
55&J161618.1+122351&199481&244.0757&12.3976&0.223&0.137&20.00&19.73&4.292&0(0)&4.07&r&1&k3i&5\\
56&J164829.7+350159&29462&252.1238&35.0330&3.526&0.203&20.32&20.16&4.075&1(1347)&1.56&D&2&rf8v&7\\
57&J171456.2+593700&215090&258.7343&59.6168&1.198&0.398&20.52&20.17&4.028&1(1406)&0.24&&6&g2v&8\\
58&J213621.4+003028&266573&324.0893&0.5079&0.602&0.082&20.54&20.09&3.2&1(1538)&0.18&G&12&wk2iii&7\\
59&J213729.8+003151&264414&324.3742&0.5310&0.263&0.043&20.97&20.80&3.630&0(0)&3.48&r&2&k01ii&5\\
60&J220814.9+015856&231039&332.0624&1.9824&0.713&0.375&19.71&19.52&3.084&0(0)&2.75&D&7&f5i&8\\
61&J220845.5+020252&235145&332.1896&2.0479&1.004&0.339&19.34&19.37&3.405&1(646)&1.38&&3&wf5v&8\\
62&J221722.2+001640&47709&334.3429&0.2780&0.088&0.019&21.28&20.61&3.366&0(0)&3.18&&2&g2i&5\\
63&J224041.7+032326&223971&340.1738&3.3906&0.262&0.091&21.15&20.45&3.348&1(1880)&3.26&&12&k3iii&6\\
\hline
\hline

	\end{longtable}
}
The structure of the table and the notation are the same as those for the table above.
%\end{landscape}

\end{center}

\end{document}